\documentclass[12pt]{article}
\usepackage{graphicx}
\usepackage{natbib} 
\usepackage{url} 

\newcommand{\blind}{0}

\usepackage{authblk}
\usepackage{amsmath}
\usepackage{amssymb}
\usepackage{tikz}
\usetikzlibrary{arrows}
\usepackage{float} 
\usepackage{booktabs}
\usepackage{algorithm}
\usepackage{algpseudocode}
\usepackage{bbm}
\usepackage{adjustbox}
\newtheorem{prop}{Proposition}
\newcommand{\bi}{\begin{itemize}}
\newcommand{\ei}{\end{itemize}}
\newcommand{\bs}[1]{\boldsymbol{#1}}
\newcommand{\cbl}[1]{{\color{black}{#1}}}

\usepackage[shortlabels]{enumitem}

\usepackage{caption}
\usetikzlibrary{bayesnet}
\newcommand*{\Scale}[2][4]{\scalebox{#1}{$#2$}}%
\usepackage{amsfonts}

\makeatletter
\g@addto@macro\normalsize{%
  \setlength\abovedisplayskip{3pt}
  \setlength\belowdisplayskip{3pt}
  \setlength\abovedisplayshortskip{3pt}
  \setlength\belowdisplayshortskip{3pt}
}
\makeatother

\begin{document}

\def\spacingset#1{\renewcommand{\baselinestretch}%
{#1}\small\normalsize} \spacingset{1}


\if0\blind
{
  \title{\bf A graphical multi-fidelity Gaussian process model, with application to emulation of heavy-ion collisions}
  \small
   \author[1]{Yi Ji}
    \author[1]{Simon Mak}
    \author[2]{Derek Soeder}
    \author[2,3]{J-F Paquet}
    \author[2]{Steffen A. Bass}
    \affil[1]{Department of Statistical Science, Duke University}
    \affil[2]{Department of Physics, Duke University}
    \affil[3]{Department of Physics and Astronomy \& Department of Mathematics, Vanderbilt University}
  \maketitle
} \fi

\if1\blind
{
  \bigskip
  \bigskip
  \bigskip
  \begin{center}
    {\LARGE\bf A graphical multi-fidelity Gaussian process model, with application to emulation of heavy-ion collisions}
\end{center}

  \medskip
} \fi

\bigskip
\begin{abstract}
With advances in scientific computing and mathematical modeling, complex scientific phenomena such as galaxy formations and rocket propulsion can now be reliably simulated. Such simulations can however be very time-intensive, requiring millions of CPU hours to perform. One solution is multi-fidelity emulation, which uses data of different fidelities to train an efficient predictive model which emulates the expensive simulator. For complex scientific problems and with careful elicitation from scientists, such multi-fidelity data may often be linked by a directed acyclic graph (DAG) representing its scientific model dependencies. We thus propose a new Graphical Multi-fidelity Gaussian Process (GMGP) model, which embeds this DAG structure (capturing scientific dependencies) within a Gaussian process framework. We show that the GMGP has desirable modeling traits via two Markov properties, and admits a scalable algorithm for recursive computation of the posterior mean and variance along \cbl{at each depth level of the DAG}. We also present a novel experimental design methodology over the DAG given an experimental budget, and propose a nonlinear extension of the GMGP via deep Gaussian processes. The advantages of the GMGP are then demonstrated via a suite of numerical experiments and an application to emulation of heavy-ion collisions, which can be used to study the conditions of matter in the Universe shortly after the Big Bang. The proposed model has broader uses in data fusion applications with graphical structure, which we further discuss.
\end{abstract}

\noindent%
\vfill

\newpage
\spacingset{1.7} 
\section{Introduction}
\label{sec:introduction}

With breakthroughs in scientific computing, computer simulations are quickly replacing physical experiments in modern scientific and engineering problems. These simulations allow scientists to better understand complex scientific problems which may be prohibitively expensive or infeasible for full-scale physical experimentation. This shift to \textit{computer} experimentation has found success in exciting applications, including cell adhesion simulation \citep{sung2020calibration} and rocket design \citep{mak2018efficient}. Such computer experiments, however, can demand a hefty price in computing resources, requiring millions of CPU hours per run. One solution is \textit{emulation} \citep{santner2019design,gramacy2020surrogates}: a handful of simulations are first run at carefully chosen design points, then an \textit{emulator} model is fit to efficiently predict the expensive computer simulator. A popular emulator is the Gaussian process (GP) model \citep{gramacy2020surrogates}, which allows for closed-form expressions for prediction and uncertainty quantification.

As systems become more realistic and complex, computer experiments also become increasingly more expensive, and thus the simulation data needed to train an accurate emulator can be difficult to generate. One way to address this is \textit{multi-fidelity emulation}. The idea is to collect data from the ``high-fidelity'' simulator, which is computationally \textit{expensive} but provides a detailed representation of the modeled science, as well as data from ``lower-fidelity'' simulators, which make simplifying assumptions on the modeled science but can be performed \textit{quickly}. One then fits a \textit{multi-fidelity} emulator using all training data to predict the output from the high-fidelity simulator. The key advantage of multi-fidelity emulation is that, by leveraging information from cheaper lower-fidelity simulations to enhance predictions for the high-fidelity model, one can train a predictive model with much fewer high-fidelity simulations and thereby lower computational costs.

The development of multi-fidelity emulators is an active research area. A popular framework is the Kennedy-O'Hagan (KO) model \citep{KOH2000}, which models a sequence of computer simulations from lowest to highest fidelity using a sequence of GP models linked by a linear autoregressive framework. The KO multi-fidelity model has been applied to a wide range of scientific problems, such as materials science \citep{PILANIA2017156} and aerodynamics \citep{lopezlopera:hal-03196283}. Further developments of the KO model include \cite{LeGratiet_2014}, which introduced a recursive computation of the predictive mean and variance given nested designs, and \cite{konomi2021bayesian}, which extended this model for non-nested designs and non-stationary responses \cbl{via a recursive Monte Carlo surrogate based on the Student $t$-process. \cite{ma2022multifidelity} explored an empirical Bayes implementation of this $t$-process using Monte Carlo expectation-maximization.} \cite{Perdikaris_2017} proposed a flexible, nonlinear extension of the KO model using deep GPs \citep{DeepGP}.



\begin{figure}[!t]
    \centering
    \includegraphics[width=0.9\linewidth]{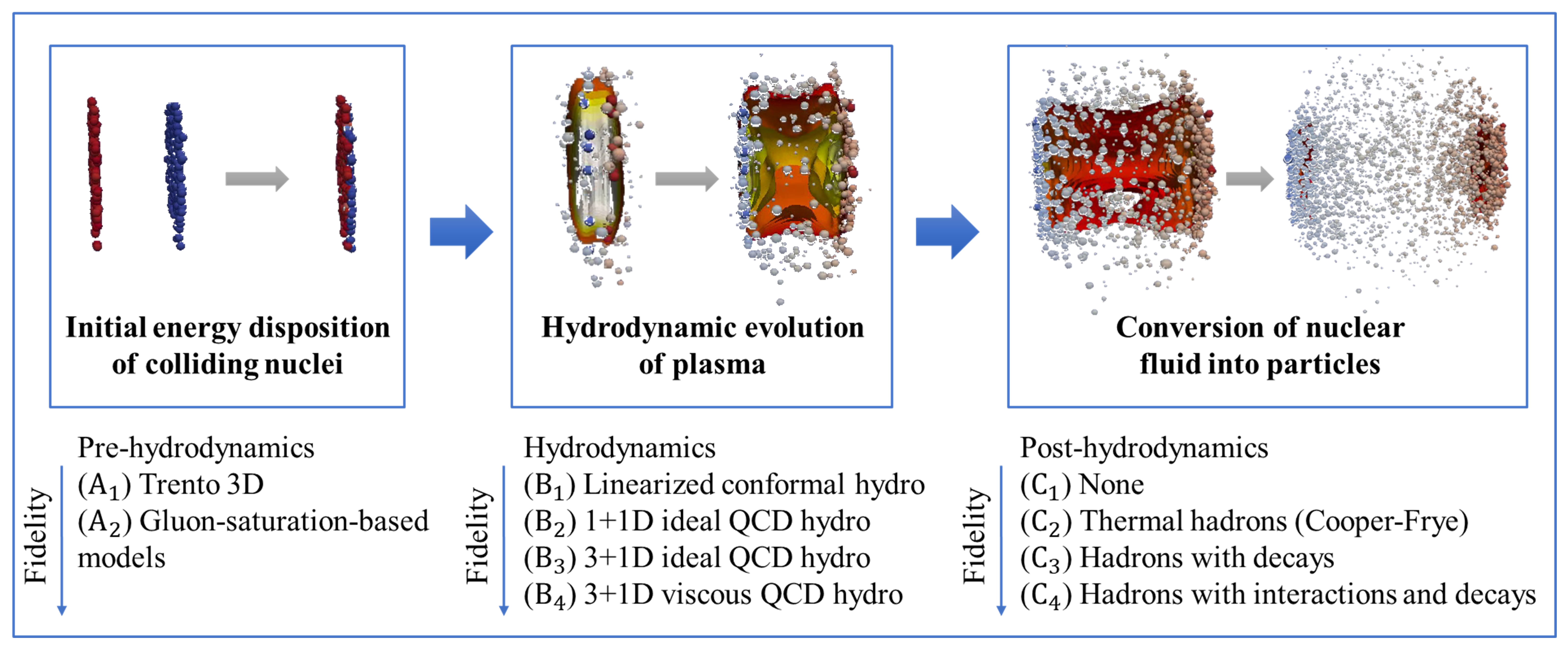}
    \caption{Visualizing the multi-stage multi-fidelity simulator for the quark-gluon plasma.}
    \label{fig:intro_flowchart}
\end{figure}

There is, however, a key limitation for the above methods: they presume the multi-fidelity data can be \textit{ranked from lowest to highest fidelity}. This may not be the case in complex scientific problems. Take, e.g., the heavy-ion collision framework in \cite{everett2021multisystem} for simulating the quark-gluon plasma, the state of nuclear matter that once filled the Universe shortly after the Big Bang, and that can now be produced and explored in collisions of heavy nuclei. This \textit{multi-stage} simulation consists of three stages, as shown in Fig. \ref{fig:intro_flowchart}. For each stage, the physicists choose one of several potential models (some more accurate but time-consuming, others less accurate but quick), resulting in many ways to perform the full plasma simulation. Here, it is difficult to rank different simulation strategies in a sequence from lowest to highest fidelity, since some may be more accurate for one stage but less accurate for another. For example, a physicist may choose the combination $A_1+B_3+C_2$ as the high-fidelity simulator $H$ for a study, and use $A_1+B_1+C_2$ and $A_1+B_2+C_1$ as two lower-fidelity approximations $L_1$ and $L_2$. To apply existing models, one may have to either (i) \textit{ignore} data from certain simulators to achieve an ordering from lowest to highest fidelity, or (ii) \textit{impose} an artificial ordering which is not justified by the science. As we shall see later, both approaches do not make full use of the underlying multi-fidelity structure, and thus may not achieve good predictive performance given a limited computing budget.

To address this, we present a new Graphical Multi-fidelity Gaussian process (GMGP) model, which utilizes a directed acyclic graph (DAG) to capture scientific dependencies between simulation models with different fidelities. This DAG structure is elicited via a careful inspection of the scientific models, which we discuss later. The GMGP embeds this DAG within a Gaussian process framework, thus allowing for a more \textit{structured} and \textit{science-driven} approach for pooling information from lower-fidelity data for high-fidelity prediction. We show that the GMGP has desirable modeling traits via two Markov properties, and admits an elegant recursive formulation that allows efficient computation of the posterior mean and variance over each depth level of the DAG. {We also present a flexible nonlinear extension of the GMGP which leverages deep GPs \citep{DeepGP}. Finally, to maximize predictive power, we propose an efficient experimental design framework for allocating multi-fidelity runs over the DAG given a computational budget.
Numerical experiments and an application to heavy-ion collisions demonstrate the improved performance of the GMGP over existing multi-fidelity models. }

While the GMGP is motivated from our nuclear physics problem, it has broad applications for other modern scientific problems. Take, e.g., the adhesion of T-cell molecules \citep{sung2020calibration}, which plays an important role in the development of immunotherapy cancer treatments \citep{harjunpaa2019cell}. The adhesion of T-cells can be modeled via two states of T-cell receptor (TCR), a resting state and an upregulated xTCR state, both of which can be modeled via computer simulations. Such simulations, however, can be very time-intensive, and a biologist may choose to run each state at varying fidelity levels, similar to the earlier heavy-ion simulator. Another application is the simulation of cosmological $N$-body problems in astrophysics, where a dark matter fluid is evolved via gravitational force. Here, the fidelity of the simulator can be controlled at different stages \citep{ho2022multifidelity}, e.g., the number of macro-particles sampled or the resolution of the spatial domain, which again yields a rich \textit{multi-stage} simulation framework. The GMGP can leverage this multi-stage structure for efficient scientific computing in astrophysics, where multi-fidelity methods have already shown great promise \citep{ho2022multifidelity}.



This work can also be viewed through the broader lens of \textit{science-driven predictive modeling}, which aims to embed known scientific principles as \textit{prior} knowledge for predictive modeling. In recent years, there has been much development on such predictive models for scientific computing, including the integration of scientific information in the form of shape constraints \citep{golchi2015monotone,wang2016estimating}, boundary constraints \citep{ding2019bdrygp}, spectral information \citep{chen2020function}, and manifold embeddings \citep{zhang2021gaussian}. Here, the GMGP integrates the DAG dependency structure of the multi-fidelity simulators (i.e., the ``science'') as prior knowledge within the GP model, which then allows for improved predictive accuracy over existing methods.

We note that the implications of the GMGP extend beyond the multi-fidelity setting into broader data fusion applications where one can elicit a graphical structure connecting data sources. One such application is networked multisensor data fusion \citep{xia2009networked}, where data are collected over different nodes on a physical sensor network. Such systems are widely used in manufacturing process monitoring and health care problems. A similar problem is distributed data fusion, where multiple agents collectively infer knowledge about a target process by sensing their local environment; this has broad applications in autonomous cars and unmanned aerial vehicles \citep{campbell2016distributed}.

It should also be noted that, for non-GP-based models, there has been some recent work \citep{2020mfnets_2,2020mfnets_1} which integrates DAG structure. These papers focus on linear subspace (e.g., polynomial-based) models, which are popular surrogate models in the applied mathematics literature. Our model has three notable distinctions. First, the GMGP makes use of GPs rather than linear subspace models, which provides greater \textit{flexibility} and \textit{robustness} in model specification \citep{gramacy2020surrogates}. Second, by leveraging error bounds for GP interpolation, we introduce a novel \textit{design} framework for multi-fidelity experiments which maximizes predictive power given a computational budget. Finally, we provide a \textit{scalable} and \textit{probabilistic} framework for modeling nonlinear linkages using deep GPs, via recursive computation of the predictive distribution at \cbl{each depth level of the DAG}. We demonstrate this in a suite of numerical experiments and an application.

The paper is organized as follows. Section 2 provides background and motivation. Section 3 presents the GMGP model, its recursive formulation, and extension for nonlinear model dependencies. Section 4 investigates experimental design approaches. Sections 5 and 6 show the effectiveness of the proposed models in a suite of numerical experiments and an application in heavy-ion collisions. Section 7 concludes the paper.

\section{Background \& Motivation}
\label{sec:background}

\subsection{Gaussian process modeling}

We first provide a brief overview of Gaussian processes (GPs). Let $\bs{x} \in \Omega \subseteq \mathbb{R}^d$ be the input parameters for the computer simulator, and $Z(\bs{x})$ be its corresponding output. A GP surrogate model places the following prior on the simulation response surface $Z$:
\[Z(\cdot)\sim \mathcal{GP}(\mu(\cdot),k(\cdot,\cdot)).\]
Here, $\mu(\bs{x}) = \mathbb{E}[Z(\bs{x})]$ is the mean function of the GP, and $k(\bs{x},\bs{x}') = \text{Cov}[Z(\bs{x}),Z(\bs{x}')]$ is its covariance function. Without prior knowledge, $\mu(\cdot)$ is typically set as constant, and $k(\cdot,\cdot)$ is chosen as the squared-exponential or Mat\'ern kernel \citep{gramacy2020surrogates}. In what follows, we assume the simulators are \textit{deterministic}
(the standard setting for computer experiments), but the proposed models extend analogously for noisy outputs.

Suppose the simulator is run at inputs $\mathcal{D}=\{\bs{x}_1, \cdots, \bs{x}_n\}$, yielding outputs $\bs{z} = \{Z(\bs{x}_1),\cdots,Z(\bs{x}_n)\}$. Conditioning on this data, the predictive distribution of $Z$ at a new input point $\bs{x}$ becomes $[Z(\bs{x})|\bs{z},\mathcal{D}] \sim \mathcal{N}(\mu_n(\bs{x}),\sigma^2_n(\bs{x}))$, with posterior mean and variance:
\begin{align}
\label{eq:meanvar}
\begin{split}
\mu_n(\bs{x}) &= \mu(\bs{x}) + \bs{k}(\bs{x},\mathcal{D})^T\bs{K}(\mathcal{D})^{-1}(\bs{z}-\bs{\mu}(\mathcal{D})),\\
\sigma^2_n(\bs{x}) &= k(\bs{x},\bs{x})-\bs{k}(\bs{x},\mathcal{D})^T\bs{K}(\mathcal{D})^{-1}\bs{k}(\bs{x},\mathcal{D}).
\end{split}
\end{align}
Here, $\bs{k}(\bs{x},\mathcal{D})=[k(\bs{x},\bs{x}_1),\cdots,k(\bs{x},\bs{x}_n)]$ is the vector of covariances, $\bs{\mu}(\mathcal{D}) = [\mu(\bs{x}_1), \cdots,\allowbreak \mu(\bs{x}_n)]$ is the vector of means, and $\bs{K}(\mathcal{D})$ is the covariance matrix for the training data. Equation \eqref{eq:meanvar} captures the key advantages of GP-based emulators: the closed-form posterior mean $\mu_n(\bs{x})$ provides an efficient \textit{emulator} of the expensive computer simulator, and the closed-form posterior variance $\sigma^2_n(\bs{x})$ quantifies its uncertainty.

\subsection{The Kennedy-O'Hagan model}

\label{sec:ko}

A popular model for multi-fidelity emulation is the Kennedy-O'Hagan model \citep{KOH2000}, which models a sequence of computer codes with increasing fidelity via a sequence of linear autoregressive GP models. Let  $\{\bs{z}_1,\bs{z}_2,\cdots,\bs{z}_T\}$ denote the data generated by $T$ levels of code sorted in increasing accuracy, where $\bs{z}_t = \{ Z_t(\bs{x}^t_i)\}_{i=1}^{n_t}$ is the data from the $t$-th code $Z_t$. The KO model assumes the multi-fidelity framework:
\begin{equation}
    \label{eqn:KOH}
      Z_t(\bs{x}) = \rho_{t-1}\cdot Z_{t-1}(\bs{x}) + \delta_t(\bs{x}), \quad Z_{t-1}(\bs{x})\perp \delta_t(\bs{x}), \quad t=2,\cdots,T.
\end{equation}
In words, Equation \eqref{eqn:KOH} presumes that, prior to data, the response surface $Z_t(\bs{x})$ can be decomposed as the lower-fidelity surface $Z_{t-1}(\bs{x})$ times a dependency parameter $\rho_{t-1}$, plus some discrepancy function $\delta_t(\bs{x})$ which models the systematic bias between the two computer simulations. For priors, the first (i.e., lowest-fidelity) surface is assigned a GP prior $Z_1(\bs{x})\sim \mathcal{GP}(\bs{h}_1(\bs{x})^T\bs{\beta}_1,\sigma_1^2r_1(\bs{x},\bs{x}'))$, and subsequent discrepancy terms are then assigned independent GP priors $\delta_t(\bs{x})\sim \mathcal{GP}(\bs{h}_t(\bs{x})^T\bs{\beta}_t,\sigma_t^2r_t(\bs{x},\bs{x}'))$. Here, the vectors $\bs{h}_1(\bs{x}), \cdots, \bs{h}_T(\bs{x})$ denote pre-defined basis functions which are used for modeling the prior mean of the GPs, and $\bs{\beta}_t$ are their associated coefficients.

\begin{figure}
\begin{minipage}{0.35\linewidth}
    \centering
    \includegraphics[width=0.9\linewidth]{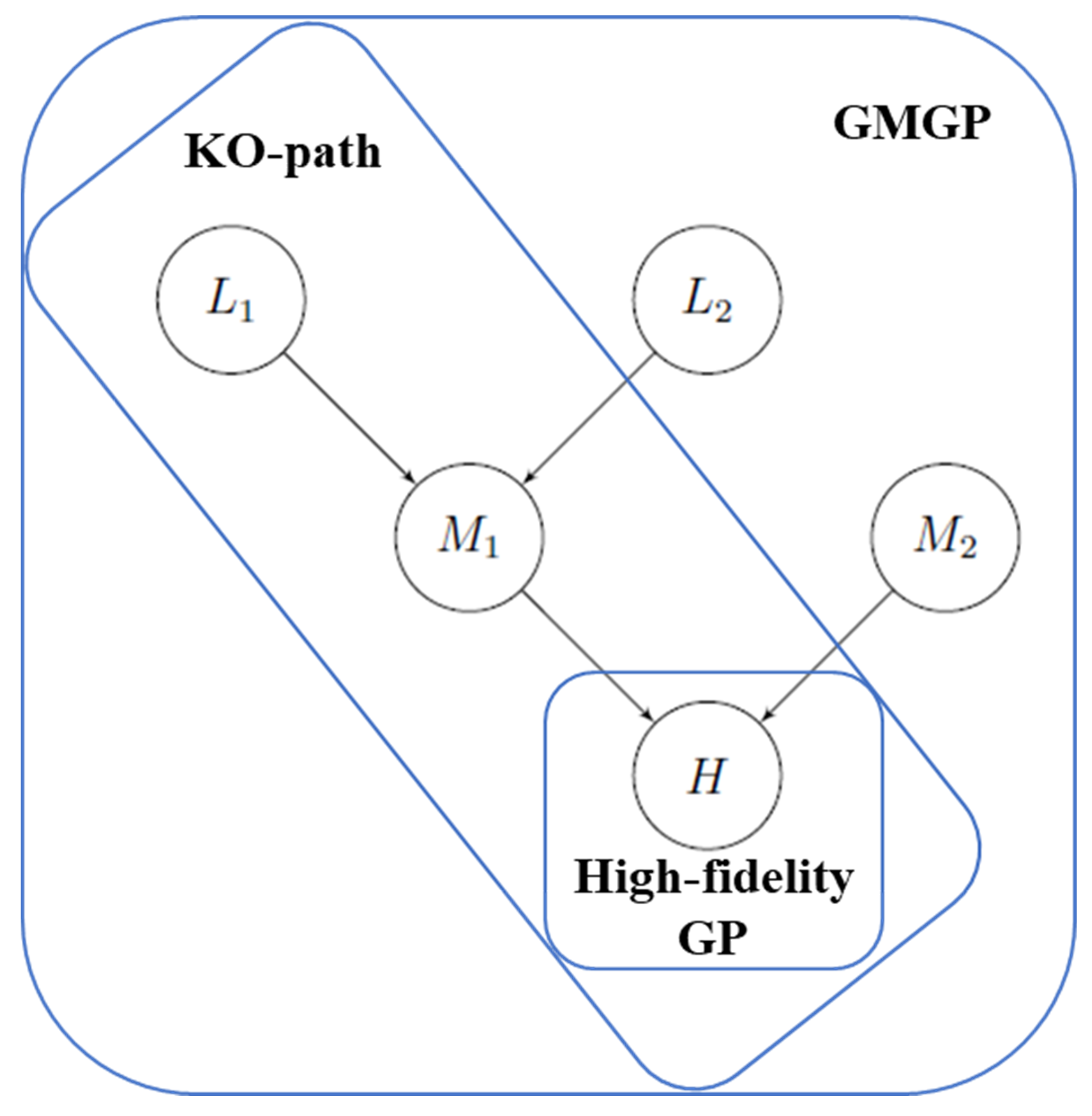}
\end{minipage}
\hspace{0.05\linewidth}
\begin{minipage}{0.6\linewidth}
    \centering
    \includegraphics[width=1\linewidth]{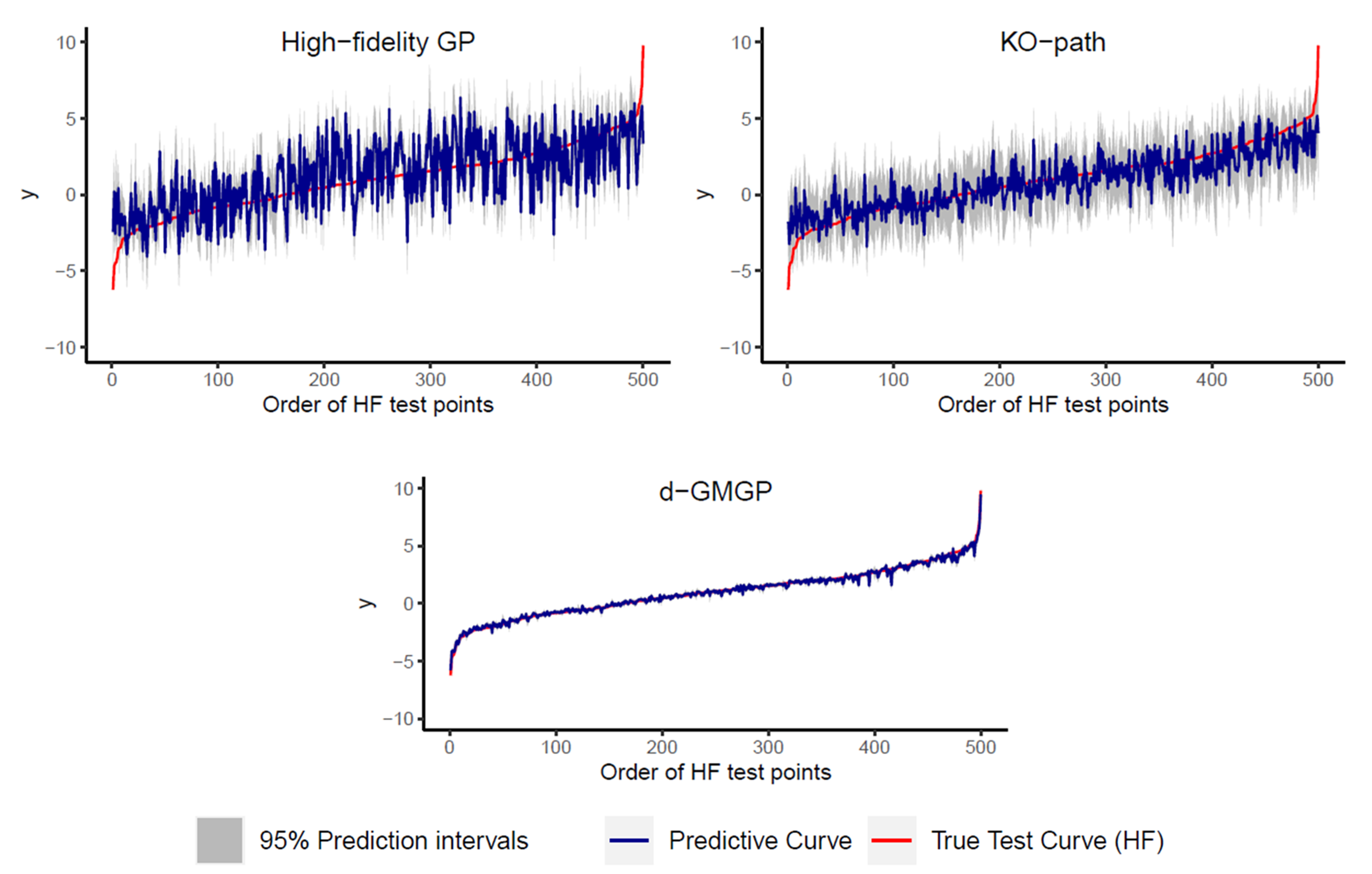}
\end{minipage}
\caption{\textbf{(Left)} The five-node DAG used in the 20-dimensional test example. \textbf{(Right)} Predictive performance of the standard high-fidelity GP model (top left), the KO-path model (top right) and the GMGP model (bottom). Red lines mark the high-fidelity (HF) outputs at test inputs, dark blue lines mark the predicted outputs, and gray bands visualize the 95\% predictive intervals.}
\label{fig:motivation}
\end{figure}


An important development in the KO model is the recursive algorithm proposed by \cite{LeGratiet_2014}, which reduces computational complexity of the KO model. The key idea is to substitute the GP prior $Z_{t-1}(\bs{x})$ in Equation \eqref{eqn:KOH} by the \textit{posterior} distribution ${Z}^*_{t-1}(\bs{x}) = [Z_{t-1}(\bs{x})|\bs{z}_1,\cdots,\bs{z}_{t-1}]$, which can be shown to follow a GP.  This recursive formulation expresses the predictive mean and variance at level $t$ as functions of the mean and variance at level $t-1$, which allows for reduced computational cost by avoiding the inversion of large covariance matrices. With a nested structure for design points, \cite{LeGratiet_2014} showed that this recursive formulation yields the same posterior predictive mean and variance as the original KO model, thus justifying the recursive approach. Further discussion will be provided later in Section \ref{sec:gmgp}.

One limitation of the KO model (and its extensions) is that it presumes the training data can be \textit{ranked} from lowest to highest fidelity. Consider a simple example where this is not the case. We take the $d=20$-d test function from \cite{WELCH1992} as the high-fidelity simulator, and generate four lower-fidelity representations (details in Section \ref{sec:simulations}). The two medium-fidelity codes ($M_1$ and $M_2$) are obtained via different simplifications on the high-fidelity code ($H$), and the two low-fidelity codes ($L_1$ and $L_2$) are obtained by different averaging operations on $M_1$. This dependence is captured by the ``multi-fidelity DAG'' in Fig. \ref{fig:motivation} (left). Here, the KO model is unable to capture this multi-fidelity structure, since the five simulators cannot be ranked in a sequence. One way to apply the KO model, which we call the ``KO-path model'', is to train it on data along the longest path $L_1$, $M_1$, $H$ (see Fig. \ref{fig:motivation} (left)); we will explore alternative ways in Section \ref{sec:simulations}.

Fig. \ref{fig:motivation} (right) shows the performance of the standard GP model trained on 25 design points on $H$, the KO-path model trained on 25 points on $H$, 50 points on $M_1$ and 75 points on $L_1$, and the proposed GMGP trained on the same data with 50 and 75 additional points on $M_2$ and $L_2$. \cbl{Further details on experimental set-up are found in Section \ref{sec:simulations}.} Compared to a standard GP, the KO-path model appears to provide slightly improved predictive performance. However, the fitted KO model still exhibits poor predictions over the input space. A natural question is whether we can improve predictive accuracy by integrating the underlying multi-fidelity DAG used for data generation (see Fig. \ref{fig:motivation} (left)). Fig. \ref{fig:motivation} (right) answers this in the affirmative: by integrating the multi-fidelity DAG within an appropriate GP model, the proposed GMGP can indeed further improve predictive performance.


\section{The Graphical Multi-fidelity Gaussian Process model}
\label{sec:gmgp}

\subsection{GMGP: model specification}

We now present the proposed GMGP model, which generalizes the KO model by embedding the underlying multi-fidelity DAG (capturing scientific dependencies between simulators) as prior information. {As noted in the Introduction, our model can be directly applied for broader data fusion settings where data sources can be linked via an underlying graphical structure. For simplicity, we defer such discussion to Section \ref{sec:concl} and focus on the multi-fidelity setting from our application.}

Let $V$ be the set of nodes representing different simulation codes, and suppose there is a \textit{root} node $T = |V| \in V$ representing the highest-fidelity simulator. Let $E$ be the set of directed edges connecting different simulation codes, where an edge $(t',t)$ is drawn \textit{only} if node $t$ is a \textit{one-step refinement} of node $t'$, i.e., $t$ is a higher-fidelity refinement of $t'$ with no intermediate codes in between. Let $\mathcal{G} = (V,E)$ be the rooted DAG for this multi-fidelity simulation framework. We will discuss later how this ``multi-fidelity DAG'' can be elicited from a careful inspection of the simulators.

Let $Z_t(\bs{x})$ be the simulation output at input $\bs{x}$ from code $t$, $t \in V$. The GMGP assumes the following modeling framework:
\begin{equation}
    \label{eqn:GMGPR_Linear}
    \begin{cases}
      Z_t(\bs{x}) = \sum_{t' \in \text{Pa}(t)}\rho_{t'}(\bs{x}) Z_{t'}(\bs{x}) + \delta_t(\bs{x}), \quad t\in \overline{{V}_{S}},\\
      Z_{t'}(\bs{x})\perp \delta_{t}(\bs{x}), \quad t'\in \text{Pa}(t).
    \end{cases}
\end{equation}
Here, $V_S \subset V$ consists of all \textit{source} nodes in $\mathcal{G}$ (i.e., nodes with an in-degree of 0), $\overline{{V}_{S}} = V \setminus V_S$ contains the remaining non-source nodes, and $\text{Pa}(t) = \{t' \in V : (t',t) \in E\}$ consists of all \textit{parent} nodes of $t \in V$ in the DAG $\mathcal{G}$. Note that source nodes represent simulations with no lower-fidelity representations, and non-source nodes represent simulations with at least one lower-fidelity representation. The function
$\rho_{t'}(\bs{x})$ captures \textit{dependencies} between the output at node $t$ and its lower-fidelity form at node $t'$. \cbl{In the absense of prior information on this dependency linking multi-fidelity models, one can instead adopt $\rho_{t}(\bs{x}) = \rho_t$; this is the specification used in later numerical experiments.}


We further assign the following GP priors on source nodes:
\begin{equation}
    \begin{cases}
        Z_t(\bs{x})\sim \mathcal{GP}(\bs{h}_t(\bs{x})^T\bs{\beta}_t,\sigma_t^2r_t(\bs{x},\bs{x'})),\quad t\in {V}_S\\
        Z_t(\bs{x}) \perp Z_{t'}(\bs{x}),\quad t,t'\in {V}_S, \quad t \neq t'.
    \end{cases}
    \label{eq:GMGPR_prior}
\end{equation}
\noindent Here, $\bs{h}_t(\bs{x})$ is a vector of basis functions for the response surface mean at node $t$, with $\bs{\beta}_t$ its coefficients. For the discrepancy term $\delta_t(\mathbf{x})$, which captures the systematic difference between $Z_t(\mathbf{x})$ and its lower-fidelity representations, we assign independent GP priors:
\begin{equation}
\delta_t(\bs{x})\sim \mathcal{GP}(\bs{h}_t(\bs{x})^T\bs{\beta}_t,\sigma_t^2r_t(\bs{x},\bs{x'})), \quad t \in \overline{{V}_{S}}.
\label{eq:GMGPR_disc}
\end{equation}
Note that this allows for different basis functions at different nodes, which can be specified based on prior information. Without such information, the bases can be set as a constant mean, i.e., $\bs{h}_t(\bs{x}) \equiv 1$, \cbl{to avoid variance inflation}. \cbl{Here, the kernels at each node employ different length-scale parameters to allow for model flexibility.} 

While the above specification may seem involved, the intuition is straight-forward. For every non-source node $t \in \overline{{V}_{S}}$, its parent nodes $\text{Pa}(t)$ contain simulations which are lower-fidelity representations of simulation $t$. Equation \eqref{eqn:GMGPR_Linear} presumes that, prior to data, $Z_t(\bs{x})$ can be decomposed as the weighted sum of its parent (lower-fidelity) simulations, plus a discrepancy term $\delta_t(\bs{x})$ to model systematic bias. The key novelty over the KO model is that, instead of pooling information in a \textit{sequence} from lowest to highest fidelity, the GMGP can integrate information over a more \textit{general} DAG structure, which better captures model dependencies between simulations of complex systems. By leveraging this graphical dependency structure guided by the underlying scientific models, the GMGP can enjoy improved predictive performance over the KO model, as we show later.

The proposition below outlines two appealing modeling properties of the GMGP:
\begin{prop}
The GMGP model satisfies the following Markov properties:
\cbl{\begin{enumerate}[(a)]
\item $Z_t(\bs{x}) \perp Z_{t'}(\bs{x}) | \{Z_j(\bs{x})\}_{j \in \textup{Pa}(t)},\quad \text{for } t' \neq t, \; t' \notin \textup{Des}(t), \; t' \notin \textup{Pa}(t)$,
\item $Z_t(\bs{x}) \perp Z_{t'}(\bs{x}') | \{Z_j(\bs{x})\}_{j \in \textup{Pa}(t)},\quad \text{for } t'\in \textup{Pa}(t),\; \bs{x}'\neq \bs{x}$, \quad \text{if $\mathcal{G}$ is a \textup{directed in-tree}.}
\end{enumerate}}
\noindent Here, $\textup{Des}(t)$ denotes the set of \textup{descendant} nodes for $t$, i.e., nodes $t'$ for which there exists a path from $t$ to $t'$.
\end{prop}

\noindent The proof is provided in the online supplement, and the formal definition of a directed in-tree is provided and justified in the next section. \cbl{We note that, without the directed in-tree structure, there may be DAGs $\mathcal{G}$ that violate property (b).}

Property $(a)$ states that, for a node $t$ with input $\bs{x}$, its output $Z_t(\bs{x})$ and the output $Z_{t'}(\bs{x})$ at another node $t'$ (where $t'$ is a non-descendant, non-parent node of $t$) are conditionally independent, given the simulation output at the parent nodes $\{Z_j(\bs{x})\}_{j \in \text{Pa}(t)}$. In other words, given knowledge of the simulator at its immediate \textit{lower-fidelity} (i.e., \textit{parent}) nodes $\text{Pa}(t)$, the output at any simulator $Z_{t'}(\bs{x})$ which is not a \textit{higher-fidelity refinement} (i.e., not a \textit{descendant}) of $t$ yields no additional information for predicting the simulator $Z_t(\bs{x})$ at node $t$. This is an intuitive modeling property if the edges in the DAG $\mathcal{G}$ indeed represent model refinements: at fixed input $\bs{x}$, simulations which are not higher-fidelity refinements of $t$ should yield little (if any) additional information on $t$ given its closest lower-fidelity representations. Property $(a)$ can be viewed as an extension of the conditional independence property for Bayesian network \citep{BNIntro}. Under a specific form for the multi-fidelity DAG $\mathcal{G}$ (which we justify in the next section), Property $(b)$ states that, conditioning on $\{Z_j(\bs{x})\}_{j \in \text{Pa}(t)}$, the simulation output $Z_t(\bs{x})$ is independent of the parent outputs $Z_{t'}(\bs{x}')$ at a different input $\bs{x}'$. In other words, given knowledge of the simulator at its immediate lower-fidelity nodes $\text{Pa}(t)$ with input $\bs{x}$, the output of such simulators at any other inputs $\bs{x}'$ yields no additional information on predicting the output $Z_t(\bs{x})$ at node $t$. This can be viewed as an extension of the Markov property in \cite{O'Hagan98amarkov}, which was used to justify the KO model.

The modeling framework \eqref{eqn:GMGPR_Linear}-\eqref{eq:GMGPR_disc} can then be used to derive the predictive distribution of the highest-fidelity simulation $Z_T(\bs{x})$. Suppose the model parameters $\Theta=\{\bs{\beta}_t$, $\sigma^2_t$, $\rho_t(\bs{x})\}_{t=1}^{T}$ are fixed (these can be estimated via maximum likelihood or a fully Bayesian approach; more on this later). Conditional on $\bs{z}^{(T)}=\{\bs{z}_1,\bs{z}_2,\cdots,\bs{z}_T\}$, where $\bs{z}_t = \{ Z_t(\bs{x}^t_i)\}_{i=1}^{n_t}$ are the observed outputs for simulator $t$, the predictive distribution for the highest-fidelity simulation at new input $\bs{x}$ is given by:
\[ [Z_T(\bs{x})|\bs{z}^{(T)},\Theta] \sim \mathcal{N}(\mu_{Z_T}(\bs{x}),\sigma^2_{Z_T}(\bs{x})),\]
where:
\begin{align} \label{eqn:GMGPR_pred}
\begin{split}
    \mu_{Z_T}(\bs{x}) &= \left[\sum_{t'\in \text{Pa}(T)}\rho_{t'}(\bs{x})\bs{h}_{t'}(\bs{x})^T\bs{\beta}_{t'}+\bs{h}_T(\bs{x})^T\bs{\beta}_T\right] + \bs{v}_T(\bs{x})^T\bs{V}_T^{-1}(\bs{z}^{(T)}-H_T\bs{\beta}), \\
    \sigma^2_{Z_T}(\bs{x}) &= v^2_{Z_T}(\bs{x}) - \bs{v}_T(\bs{x})^T\bs{V}_T^{-1}\bs{v}_T(\bs{x}).
\end{split}
\end{align}
Here, $\bs{v}_T(\bs{x})=\text{Cov}(Z_T(\bs{x}),\bs{z}^{(T)})$ is the covariance vector of the new observation $Z_T(\bs{x})$ and data $\bs{z}^{(T)}$, $\bs{V}_T = \text{Var}(\bs{z}^{(T)})$ is the covariance matrix of $\bs{z}^{(T)}$, $v^2_{Z_T}(\bs{x}) = \text{Var}(Z_T(\bs{x}))$ is the prior variance of $Z_T(\bs{x})$, $H_T$ is the matrix of basis functions and $\bs{\beta}=\{\bs{\beta}_1,\cdots,\bs{\beta}_T\}$ are the coefficients such that $H_T\bs{\beta}$ yields the vector of prior means for $\bs{z}^{(T)}$ from \eqref{eqn:GMGPR_Linear}.

While Equation \eqref{eqn:GMGPR_pred} provides closed-form expressions for the predictive mean and variance, such expressions can be unwieldy to compute due to the inverse of the large matrix $\bs{V}_T$, which has dimensions $\sum_{t=1}^{T} n_t \times \sum_{t=1}^{T} n_t$. \cbl{Motivated by \cite{LeGratiet_2014}, we consider below a recursive formulation for the GMGP that performs model training at each depth level of the DAG, thus enabling scalable predictions.}


\subsection{r-GMGP: recursive formulation}

\cbl{The key idea for the r-GMGP is to recursively perform model training and prediction at each depth level of $\mathcal{G}$, beginning with source (i.e., depth 0) nodes, then continuing for higher depth nodes until the highest-fidelity node is reached. } 
This is facilitated by the following recursive computation of the predictive distribution, which we later show yields the desired predictive mean and variance from the GMGP \eqref{eqn:GMGPR_Linear} under certain conditions:
\begin{equation}
    \label{eqn:GMGPR_Linear_rec}
    \begin{cases}
      Z_t(\bs{x}) = \sum_{t'\in \text{Pa}(t)}\rho_{t'}(\bs{x}) Z^*_{t'}(\bs{x}) + \delta_t(\bs{x}),\quad t\in \overline{{V}_{S}},\\
      Z^*_{t'}(\bs{x})\perp \delta_t(\bs{x}),\quad t'\in \text{Pa}(t).
    \end{cases}
\end{equation}
Here, $Z^*_{t'}(\cdot)=[Z_{t'}(\cdot)| \{\bs{z}_m\}_{m \in \text{Anc}(t')},\bs{z}_{t'},\cbl{\Theta_{t'}}]$ is the \textit{posterior} distribution at node $t'$, conditional on data from both $t'$ and its ancestor nodes $\text{Anc}(t')$, i.e., nodes $u$ where there exists a path from $u$ to $t'$. \cbl{The exact expression for $Z^*_{t'}(\cdot)$ is given in Equation \eqref{eqn:rec_pred} below, with $T$ replaced by the current node.} The same GP priors \eqref{eq:GMGPR_disc} are assigned for the discrepancy terms $\delta_t(\bs{x})$. The key difference between the recursive GMGP (r-GMGP) formulation \eqref{eqn:GMGPR_Linear_rec} and the GMGP \eqref{eqn:GMGPR_Linear} is that, in place of $Z_{t'}(\bs{x})$ (the \textit{prior} of the parent simulation) for the GMGP, the r-GMGP uses $Z^*_{t'}(\bs{x})$, the \textit{posterior} of the parent simulation given data. 

Under \eqref{eqn:GMGPR_Linear_rec}, the posterior distribution of the highest-fidelity simulation $Z_T$ becomes:
\[ [Z_T(\bs{x})|\bs{z}^{(T)},\Theta] \sim \mathcal{N}(m_{Z_T}(\bs{x}),s^2_{Z_T}(\bs{x})),\]
with the posterior mean $m_{Z_T}(\bs{x})$ and variance $s^2_{Z_T}(\bs{x})$:
\small
\begin{align}\label{eqn:rec_pred}
\small
\begin{split}
    m_{Z_T}(\bs{x}) &= \sum_{t'\in \text{Pa}(T)}\rho_{t'}(\bs{x}) m_{Z_{t'}}(\bs{x})+\bs{h}_{T}(\bs{x})^T\bs{\beta}_T
    \\
    & \quad \quad \quad  + \bs{r}_{T}(\bs{x},\mathcal{D}_T)^T\bs{R}_{T}(\mathcal{D}_T)^{-1}\left[\bs{z}_T -\sum_{t'\in\text{Pa}(T)}\bs{\rho}_{t'}(\mathcal{D}_{T})\odot \bs{z}_{t'}(\mathcal{D}_{T}) - \bs{h}_{T}(\mathcal{D}_{T})^T\bs{\beta}_{T}\right], \\
    s^2_{Z_T}(\bs{x}) &= \sum_{t'\in \text{Pa}(T)}\rho^2_{t'}(\bs{x})s^2_{Z_{t'}}(\bs{x}) + \sigma^2_{T}\left[1-\bs{r}_{T}(\bs{x},\mathcal{D}_T)^T\bs{R}_{T}(\mathcal{D}_T)^{-1}\bs{r}_{T}(\bs{x},\mathcal{D}_T)\right].
\end{split}
\end{align}
\normalsize
Here, $\odot$ denotes the Hadamard (entrywise) product, $\mathcal{D}_T = \{\bs{x}_i^T\}_{i=1}^{n_T}$ is the set of design points at node $T$, $\bs{r}_T(\bs{x},\mathcal{D}_T)$ is the correlation vector between $Z_T(\bs{x})$ and $\bs{z}_T$, and $\bs{R}_T(\mathcal{D}_T)$ is the correlation matrix of $\bs{z}_T$. 
Equation \eqref{eqn:rec_pred} provides closed-form expressions for the predictive distribution of the highest-fidelity simulation $T$, which depend on only terms related to the current node $T$ and its parent nodes $\text{Pa}(T)$. Indeed, Equation \eqref{eqn:rec_pred} holds for all non-source nodes $t \in \overline{V_S}$, with closed-form expressions depending on only node $t$ and its parent nodes $\text{Pa}(t)$. Thus, the desired predictive distribution $[Z_t(\bs{x})|\bs{z}^{(T)},\Theta]$ can be efficiently evaluated, by \textit{recursively} computing the posterior mean and variance using \eqref{eqn:rec_pred} \cbl{at each depth level of} $\mathcal{G}$, starting from its leaf nodes to its root. We show later that the predictive equations \eqref{eqn:rec_pred} for r-GMGP are precisely the desired predictive equations \eqref{eqn:GMGPR_pred} for the GMGP model under certain conditions.

This recursive approach can yield significant computational savings over a naive evaluation of the original predictive equations \eqref{eqn:GMGPR_pred}, since it breaks up the inversion of the \textit{large} covariance matrix $\bs{V}_T$ into inversions of \textit{smaller} matrices $\bs{R}_t(\mathcal{D}_t)$ \cbl{at each depth level of $\mathcal{G}$}. More precisely, with $n_t$ denoting the sample size at node $t \in V$, this reduces the computational cost of $\mathcal{O}((\sum_{t\in V}n_t)^3)$ for the original predictive equations \eqref{eqn:GMGPR_pred} to a cost of $\mathcal{O}(\sum_{t\in V}n_t^3)$ for the recursive approach in \eqref{eqn:rec_pred}. Furthermore, if the inverse at each level of $\mathcal{G}$ can be performed simultaneously via \textit{distributed} computing, this computational cost can be further reduced to $\mathcal{O}(D \max_t(n_t)^3)$, where $D \leq T$ is the depth of the rooted graph $\mathcal{G}$. When the sample sizes are moderately large at most nodes (which is typically the case for low-fidelity simulations), this recursive computation of the posterior mean and variance \cbl{at each depth level} can yield large computational savings.

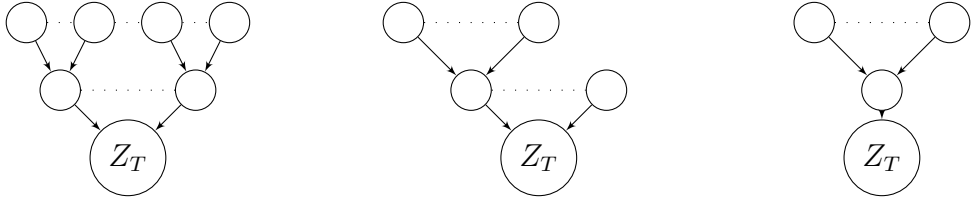
\begin{figure}[!t]
\centering
\begin{minipage}{0.3\textwidth}
\centering
\begin{tikzpicture}
    \tikzset{vertex/.style = {shape=circle,draw,minimum size=1.3em}}
    \tikzset{edge/.style = {->,> = latex'}}
    \node[vertex] (a) at  (0,0) {};
    \node[vertex] (b) at  (0.9,0) {};
    \node[vertex] (c) at  (1.8,0) {};
    \node[vertex] (d) at  (2.7,0) {};
    \node[vertex] (e) at  (0.45,-0.9) {};
    \node[vertex] (f) at  (2.25,-0.9) {};
    \node[vertex] (g) at  (1.35,-1.8) {$Z_T$};
    \draw[edge] (a) to (e);
    \draw[edge] (b) to (e);
    \draw[edge] (c) to (f);
    \draw[edge] (d) to (f);
    \draw[edge] (e) to (g);
    \draw[edge] (f) to (g);
    \draw[loosely dotted] (a) to (b);
    \draw[loosely dotted] (b) to (c);
    \draw[loosely dotted] (c) to (d);
    \draw[loosely dotted] (e) to (f);
\end{tikzpicture}
\end{minipage}
\begin{minipage}{0.3\textwidth}
    \centering
    \begin{tikzpicture}
    \tikzset{vertex/.style = {shape=circle,draw,minimum size=1.3em}}
    \tikzset{edge/.style = {->,> = latex'}}
    \node[vertex] (a) at  (-0.45,0) {};
    \node[vertex] (c) at  (1.35,0) {};
    \node[vertex] (e) at  (0.45,-0.9) {};
    \node[vertex] (f) at  (2.25,-0.9) {};
    \node[vertex] (g) at  (1.35,-1.8) {$Z_T$};
    \draw[edge] (a) to (e);
    \draw[edge] (c) to (e);
    \draw[edge] (e) to (g);
    \draw[edge] (f) to (g);
    \draw[loosely dotted] (a) to (c);
    \draw[loosely dotted] (e) to (f);
    \end{tikzpicture}
\end{minipage}
\begin{minipage}{0.3\textwidth}
    \centering
    \begin{tikzpicture}
    \tikzset{vertex/.style = {shape=circle,draw,minimum size=1.3em}}
    \tikzset{edge/.style = {->,> = latex'}}
    \node[vertex] (a) at  (0,0) {};
    \node[vertex] (b) at  (1.8,0) {};
    \node[vertex] (c) at  (0.9,-0.9) {};
    \node[vertex] (d) at  (0.9,-1.8) {$Z_T$};
    \draw[edge] (a) to (c);
    \draw[edge] (b) to (c);
    \draw[edge] (c) to (d);
    \draw[loosely dotted] (a) to (b);
    \end{tikzpicture}
\end{minipage}
\caption{Examples of in-trees, with $Z_T$ denoting the highest-fidelity simulator at the root.}
\label{fig:intree} 
\end{figure}

We now return to the important question of whether the r-GMGP predictive equations \eqref{eqn:rec_pred} are indeed the same as the desired predictive equations \eqref{eqn:GMGPR_pred} for GMGP. To show this equivalence, we require a specific structure of the multi-fidelity DAG $\mathcal{G}$. Suppose $\mathcal{G}$ is a directed \textit{in-tree} \citep{mehlhorn2008algorithms}, defined as a rooted tree for which, at any node $t \in V$, there is exactly \textit{one} path going from node $t$ to the root node $T$ (representing the highest-fidelity simulator $Z_T$). In-trees are also known as \textit{anti-arborescence} trees \citep{bernhard2011combinatorial} in graph theory. Fig. \ref{fig:intree} shows several examples of directed in-trees.
Under such an assumption on $\mathcal{G}$, the following proposition shows that the r-GMGP indeed yields the desired posterior predictive mean and variance for the highest-fidelity simulation $Z_T$ for the GMGP model:
\begin{prop}
Suppose the rooted multi-fidelity DAG $\mathcal{G} = (V,E)$ is an in-tree. Further suppose (i) the observations are noise-free, and (ii) the design points are nested over $\mathcal{G}$, such that for any node $t \in V$, its design set $\mathcal{D}_{t}$ is a subset of the designs at all parent nodes $\mathcal{D}_{t'},\ t' \in \textup{Pa}(t)$. Then, conditional on the parameters $\Theta=\{\bs{\beta}_t$, $\sigma^2_t$, $\rho_t(\bs{x})\}_{t=1}^{T}$ of both models, the posterior predictive mean and variance from GMGP and r-GMGP are the same, i.e., $\mu_{Z_T}(\bs{x}) = m_{Z_T}(\bs{x})$ and $\sigma^2_{Z_T}(\bs{x}) = s^2_{Z_T}(\bs{x})$, where $\mu_{Z_T}(\bs{x})$ and $\sigma^2_{Z_T}(\bs{x})$ are the GMGP posterior mean and variance in \eqref{eqn:GMGPR_pred}, and $m_{Z_T}(\bs{x})$ and $s^2_{Z_T}(\bs{x})$ are the r-GMGP posterior mean and variance in \eqref{eqn:rec_pred}.
\end{prop}
\noindent The proof \cbl{(by induction)} is provided in the online supplement. This proposition shows that the recursive GMGP formulation indeed yields the same predictive mean and variance as the GMGP model when the multi-fidelity graph $\mathcal{G}$ forms an in-tree, thus justifying the computational savings from r-GMGP. The assumption of $\mathcal{G}$ being an in-tree implies that, for any simulator (i.e., node), there exists exactly one path in the employed simulation framework along which this model can be refined to the highest-fidelity simulator (i.e., root node). For example, the three heavy-ion collision models in the Introduction form a 3-node in-tree (see Fig. \ref{fig:1dexperiment} (left)). In practice, such a property can often be satisfied via a careful \textit{choice} of lower-fidelity simulators to run for training the multi-fidelity emulator. \cbl{In cases where the simulators cannot be selected and do not form an in-tree, the original GMGP equations \eqref{eqn:GMGPR_pred} may be used for prediction, albeit at higher costs from larger matrix inversions.}

For inference on model parameters, we employ a straight-forward extension of the maximum likelihood approach in \cite{LeGratiet_2014}, which accounts for uncertainties in regression and dependency parameters within a universal co-kriging framework. Details can be found in \cite{LeGratiet_2014}. In particular, our implementation of r-GMGP is built upon the R package \texttt{MuFiCokriging} \citep{R_MFCK} for this paper, which is available on CRAN. If fully Bayesian inference is desired on such parameters, one can adapt the Monte Carlo approach in \cite{konomi2021bayesian}. 

\cbl{Finally, we note that while the above formulation presumes independence over source nodes, there may be situations where prior information suggests some correlation may be preferable between these nodes. In such cases, one may adopt correlated GP priors over source nodes, then use the original GMGP without recursive updates.}

\subsection{d-GMGP: nonlinear extension}
\label{sec:dGMGP}

One potential limitation of the GMGP is that it presumes linear dependencies between the simulation codes over the DAG. Given enough training data, it may be preferable to consider a more sophisticated emulator model that accounts for \textit{nonlinear} dependencies between nodes on the multi-fidelity DAG. \cbl{There are multiple ways for modeling this, including the deep GP approach in \cite{Perdikaris_2017} and the binary tree partition method in \cite{konomi2021bayesian}. Below, we adapt the deep GP approach for two reasons: (i) the output observables in our high-energy physics application are known to be quite smooth \citep{everett2021multisystem}, thus the smoother deep GP approach is preferable to binary tree partitions; (ii) our extension avoids the need for MCMC sampling in approximating the predictive distribution.}


Similar to before, let us assume the multi-fidelity DAG $\mathcal{G}$ is a directed in-tree, with design points nested over $\mathcal{G}$. The d-GMGP model at non-source nodes is formulated as:
\begin{equation}\label{eqn:GMGPR_nonlinear}
    \begin{cases}
      Z_t(\bs{x}) = f_t(\{Z_{t'}(\bs{x}):t'\in \text{Pa}(t)\} \cup \bs{x})+\delta_t(\bs{x}), \quad t \in \overline{{V}_{S}},\\
      f_t(\cdot) \perp \delta_t(\cdot).
    \end{cases}
\end{equation}
Here, we again assign independent GP priors for the discrepancies $\delta_t(\bs{x})\sim \mathcal{GP}(0,\sigma_t^2r_t(\bs{x},\bs{x}'))$, with similar independent GP priors on source nodes:
\begin{equation}
    \begin{cases}
        Z_t(\bs{x})\sim \mathcal{GP}(0,\sigma_t^2r_t(\bs{x},\bs{x'})),\quad t\in {V}_S,\\
        Z_t(\bs{x}) \perp Z_{t'}(\bs{x}),\quad t,t'\in {V}_S, \quad t \neq t'.
    \end{cases}
\end{equation}
\noindent The key difference between this new model and the GMGP model \eqref{eqn:GMGPR_Linear} is how the lower-fidelity (parent) nodes are integrated for higher-fidelity models. Instead of a weighted sum of the parent codes $\sum_{t' \in \text{Pa}(t)}\rho_{t'}(\bs{x}) Z_{t'}(\bs{x})$, the d-GMGP model allows for a more \textit{general} nonlinear transformation $f_t(\cdot)$ of the parent codes $\{Z_{t}(\bs{x}):t\in \text{Pa}(t)\}$ as well as the control parameters $\bs{x}$.

Since the transformation $f_t$ is unknown in practice, one approach is to assign to it an independent zero-mean GP prior. We can combine the GP priors on $f_t$ and $\delta_t$ with Equation \eqref{eqn:GMGPR_nonlinear} to obtain the general specification for $Z_t(\bs{x})$ for non-source node $t \in \overline{V_S}$:
\begin{equation}
Z_t(\bs{x}) = g_t(\{Z_{t'}(\bs{x}):t'\in \text{Pa}(t)\} \cup \bs{x}) \sim \mathcal{GP}(0,K_t([\bs{x},\bs{z}],[\bs{x}',\bs{z}'])),\quad t \in \overline{V_S},
\label{eq:dgmgpkern}
\end{equation}
where $\bs{z} = \{Z_{t'}(\bs{x})\}_{t'\in \text{Pa}(t)}$ and $\bs{z}' = \{Z_{t'}(\bs{x}')\}_{t'\in \text{Pa}(t)}$.
Note that the kernel $K_t$ involves both the control parameters $\bs{x}$ and the simulation outputs $Z_{t'}(\bs{x})$ from lower-fidelity (parent) nodes $t' \in \text{Pa}(t)$. Viewed this way, the proposed d-GMGP model can be seen as an extension of the deep GP model (see, e.g., \citealp{DeepGP}), where the GP outputs at each node are linked by the multi-fidelity DAG elicted from model dependencies (i.e., the ``science''). Compared to a full-blown deep GP model, which typically requires the estimation of thousands of variational parameters, the d-GMGP model \eqref{eq:dgmgpkern} with the kernel choice below requires much fewer parameters for estimation, all the while providing the desired nonlinear dependency over the graph.

For the kernel $K_t$ in \eqref{eq:dgmgpkern}, one specification we found quite effective is the following:
\begin{equation}
K_t([\bs{x},\bs{z}],[\bs{x}',\bs{z}']) = K_{\rm SE, \cbl{\rho}}(\bs{x},\bs{x}') \left[K_{\rm LIN}(\bs{z},\bs{z}') + K_{\rm SE}(\bs{z},\bs{z}')\right]+K_{\rm SE, \cbl{\delta}}(\bs{x},\bs{x}'),\quad t \in \overline{V_S}.
\label{eqn:d-GMGP_kernel}
\end{equation}
Here, $K_{\rm LIN}(\bs{z},\bs{z}')= \sigma^2\bs{z}^T\bs{z}'$ is a linear kernel, $\sigma^2$ is a variance parameter, and \cbl{$K_{\rm SE}(\bs{z},\bs{z}')$, $K_{\rm{SE},\rho}(\bs{x},\bs{x}')$ and $K_{\rm{SE},\delta}(\bs{x},\bs{x}')$ are separate anisotropic squared-exponential kernels}. This is motivated by the kernel choice for the deep multi-fidelity GP models in \cite{Perdikaris_2017} and \cite{cutajar2019deep}. The intuition behind \eqref{eqn:d-GMGP_kernel} is that it captures both linear and nonlinear dependencies between outputs, as well as correlations between input parameters. When $K_{\rm SE}(\bs{z},\bs{z}') = 0$, this kernel reduces to a form similar to the r-GMGP, with a probabilistic and non-parametric form for $\rho(\bs{x})$. \cbl{In total, the d-GMGP with kernel \eqref{eqn:d-GMGP_kernel} requires $(1+d)|V_S|+\sum_{t\in \overline{V_S}}(4+2d+|\text{Pa}(t)|)$ hyperparameters to estimate (via maximum likelihood), which is much fewer than that needed for a full-scale deep GP; this is primarily due to the above recursive formulation. Such a deep model, however, requires the estimation of more parameters compared to the r-GMGP and thus requires more computation for inference and prediction; it should therefore be used only when one has prior information on nonlinear dependencies. When model outputs are known to be highly non-stationary and/or discontinuous, the binary tree partition approach in \cite{konomi2021bayesian} may offer an appealing alternative for computational efficiency.}

As before, a recursive formulation can be adopted for efficient fitting of the d-GMGP. \cbl{The idea is again to recursively perform model training and prediction at each depth level of $\mathcal{G}$.} This is achieved by replacing the GP priors $Z_{t'}(\cdot)$, $t' \in \text{Pa}(t)$ in Equation \eqref{eqn:GMGPR_nonlinear} by the GP posteriors
$Z^*_{t'}(\cdot)=[Z_{t'}(\cdot)| \{\bs{z}_m\}_{m \in \text{Anc}(t')},\bs{z}_{t'},\cbl{\Theta_{t'}}]$, $t' \in \text{Pa}(t)$. The desired posterior on the highest-fidelity node $Z_T(\cdot)$ can then be computed recursively \cbl{at each depth level of} $\mathcal{G}$, starting from its leaf nodes to its root, as was done for the r-GMGP. Unlike the r-GMGP, however, the predictive distribution for the highest-fidelity node $Z_T(\cdot)$ is no longer Gaussian. This recursive formulation allows for efficient approximation of the desired predictive distribution, by propagating the posterior uncertainty at each level using Monte Carlo. Specifically, the posterior distribution at a non-source node $t$ can be evaluated by:
\begin{equation}
 [Z_t^*(\bs{x})]=\int [Z_t(\bs{x})|\bs{z}_t,\bs{z^*}] \prod_{t'\in \text{Pa}(t)}[Z_{t'}^*(\bs{x})] \; d\bs{z}^*.
\end{equation}
Here, $\bs{z}_t$ are the observed data on node $t$, and $\bs{z^*} = \{Z^*_{t'}(\bs{x})\}_{t'\in \text{Pa}(t)}$ are the (unknown) outputs on parent nodes at parameters $\bs{x}$. This can be estimated via Monte Carlo integration on the posterior of parent node outputs $[Z_{t'}^*(\bs{x})]$, $t' \in \text{Pa}(t)$. This procedure can then be repeated recursively \cbl{at each depth level of $\mathcal{G}$} to provide efficient computation of the predictive distribution for the highest-fidelity node $Z_T(\cdot)$. Compared to full-scale deep GP model, this recursive approach provides a scalable way for propagating predictions and uncertainties to the root node, without the need for complex variational approximations. The full algorithm for d-GMGP prediction is provided in the online supplement.

\cbl{For the d-GMGP, the computational cost for hyperparameter estimation using maximum likelihood is $\mathcal{O}(\sum_{t\in V}n_t^3)$ per objective evaluation, which greatly speeds up the $\mathcal{O}((\sum_{t\in V}n_t)^3)$ cost for standard GP via its recursive formulation. Its prediction then requires \textit{sampling} from the posterior distribution of each parent node model, then \textit{propagating} these as inputs of each child node until we reach the root of the tree. Here, the posterior samples needed to achieve a desired accuracy for highest-fidelity prediction can grow exponentially with both the input dimensions and the size of the tree. One solution (which we adopt) is to employ a Gaussian approximation of the posterior predictive distribution at each node, then recursively compute the posterior means and variances over the nodes in DAG $\mathcal{G}$. The latter step can be performed via the closed-form expressions in \cite{girard2003gaussian}, and further details of this can be found in \cite{Perdikaris_2017}.}

In choosing between the original GMGP model (which models linear dependencies) or the above d-GMGP model, we have found that with careful elicitation from scientists, there is often prior knowledge on scientific model dependencies which can help guide this choice. For example, in computational fluid dynamics (see, e.g., \citealp{wang2016swirling}), the dependencies between the high-fidelity direct numerical simulation \citep{pope2000turbulent} and the lower-fidelity Reynolds-averaged Navier Stokes simulation \citep{catalano2003evaluation} is known to be highly nonlinear, and captures complex eddies and vortices in turbulent fluid flow. For this case, the d-GMGP should be thus used instead of the original GMGP model. In the absence of such prior information, one can make use of standard model selection techniques (e.g., AIC or BIC) to select the better predictive model from data.

\section{Experimental Design}
\label{sec:Design}

Given that the motivation behind the GMGP model is to maximize predictive performance given a computational budget, its \textit{experimental design} is of crucial importance. This procedure can be split into two steps: (i) the \textit{design} of training set $\mathcal{D}_t$ at each node $t \in V$ given \textit{fixed} sample sizes $n_t = |\mathcal{D}_t|$, and (ii) the \textit{allocation} of sample sizes $n_t$ at each node $t \in V$ given a \textit{fixed} computational budget. For simplicity, we investigate this for the GMGP model, but such designs can naturally be adapted for the more complex d-GMGP model. As before, we assume the underlying DAG is an in-tree.

\subsection{Design given fixed sample sizes}
\label{sec:bfs}

\begin{figure}[!t]
    \centering
    \includegraphics[width=0.7\textwidth]{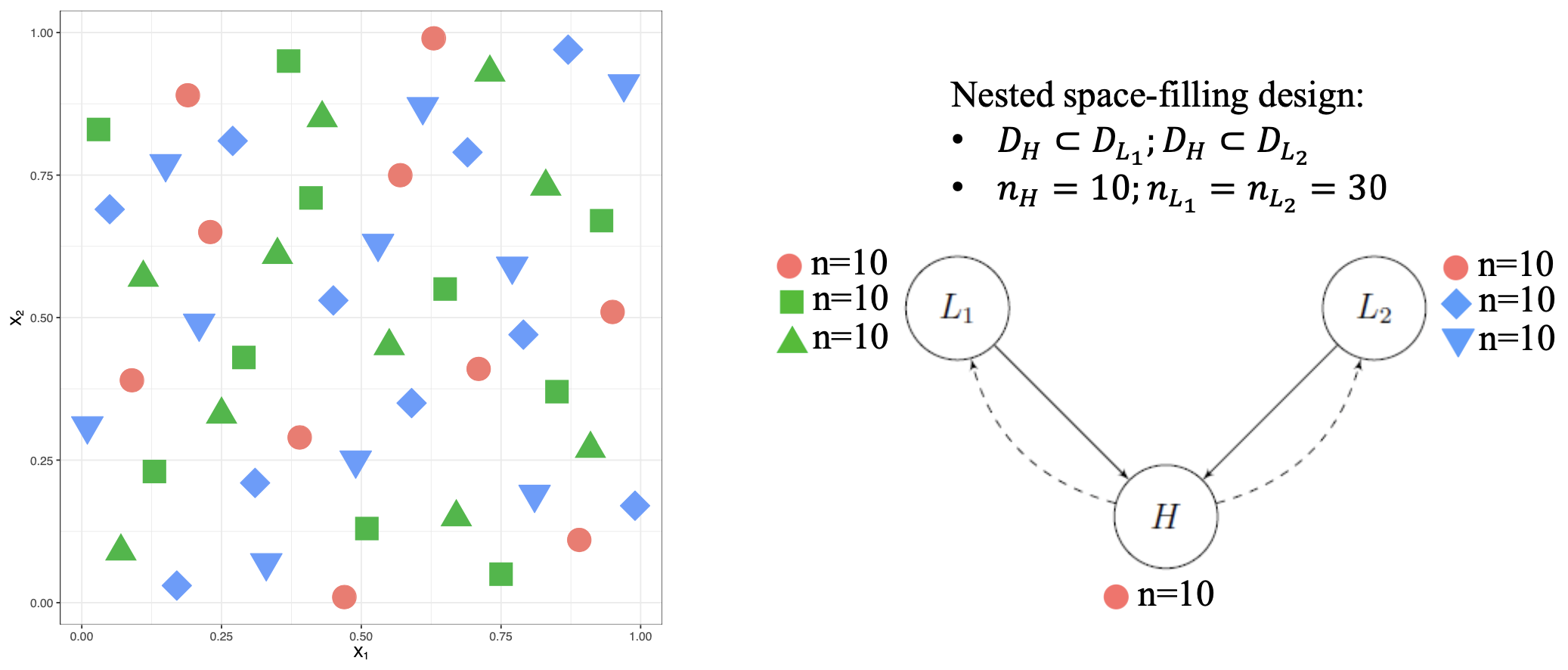}
    \caption{Visualizing our nested BFS design on a $2$-d space. (\textbf{Left}) The base maximin SLHD design with 5 slices and 10 design points per slice. (\textbf{Right}) BFS allocation of the SLHD design over a three-node DAG. The first slice (red) is allocated to all three nodes, then subsequent slices (green and blue) are used on the lower-fidelity nodes $L_1$ and $L_2$.}
    \label{fig:BFSN}
\end{figure}

Consider first step (i). From Proposition 2, an appealing design property is the \textit{nested} nature of design points over the graph $\mathcal{G}$, i.e., for any node $t \in V$, the design set $\mathcal{D}_t$ is a subset of the designs $\mathcal{D}_{t'}$ at any parent node $t' \in \text{Pa}(t)$. For GP modeling, the \textit{space-filling} property of design points \citep{santner2019design} -- its uniformity over the prediction space -- is also known to be crucial for improving predictive performance. Different notions of space-fillingness have been explored in the literature, including maximin \citep{johnson1990minimax,morris1995exploratory} and minimax designs \citep{johnson1990minimax,mak2018minimax}. We will incorporate these two properties in the design procedure below.

\begin{algorithm}[!t]
\caption{Nested BFS Design}\label{alg:DAGE}
\textbf{Input:} DAG with $T=|V|$ nodes; desired sample sizes $\{n_t:\ t\in V\}$. Note that all $n_t$'s should be multiples of $n_T$.\\
\textbf{Output:} Design set $\mathcal{D}_t$ for each node $t \in V$.
\begin{algorithmic}[1]
    \State Generate a maximin SLHD design \citep{Ba2015SLHD} with $M=\sum_{t=1}^T{n_t}/{n_T}$ slices, with each slice containing $n_T$ design points.
    \State Assign the design points in slices $\{\sum_{j=1}^{t-1}{n_j}/{n_T}+1,\cdots,\sum_{j=1}^{t}{n_j}/{n_T}\}$ (denoted as $S_t$) to node $t$, $t = 1, \cdots, T$.
    \State For each node $t$, set $\mathcal{D}_t=S_t \cup\ 
    \left\{\cup_{t'\in \text{Des}(t)}S_{t'}\right\}$ for  $t=1,\cdots,T$.
\end{algorithmic}
\end{algorithm}

Given sample sizes $n_1, \cdots, n_T$, we propose a nested experimental design over the DAG $\mathcal{G}$. We make use of the maximin sliced Latin hypercube design (maximin SLHD, \citealp{Ba2015SLHD}; see also \citealp{qian2012sliced}), which provides design points in equal \textit{slices} (or batches), such that the design points \textit{within} each slice are space-filling, and the design points \textit{between} slices are also well spaced-out. Fig. \ref{fig:BFSN} (left) shows an SLHD in $d=2$ dimensions. With this SLHD in hand, we then employ a \textit{bottom-up} approach to allocate design points over $\mathcal{G}$. We first allocate one slice in the SLHD for the \textit{highest-fidelity} simulator $T$ (i.e., the root node at the bottom of $\mathcal{G}$), then use the remaining slices to fill out design points on subsequent nodes in a \textit{breadth-first-traversal} (BFS) of $\mathcal{G}$. For the latter step, the design points at each node $t \in V$ are obtained by concatenating the current SLHD slice(s) with design points on its descendant nodes $\text{Des}(t)$; this ensures the design is nested over the DAG $\mathcal{G}$. Fig. \ref{fig:BFSN} (right) visualizes this nested BFS design procedure. Intuitively, the use of maximin SLHDs allows one to ``maximize'' the information obtained over different nodes for predicting the high-fidelity response surface (we provide a more formal discussion of this in the next subsection). Algorithm 1 provides the detailed steps.

To implement this nested BFS design, however, there are certain conditions which need to hold for the sample sizes $n_1, \cdots, n_T$. The first condition is that $n_{t'} \geq n_t$ if edge $(t',t) \in E$, i.e., a lower-fidelity node should always have as many sample points as its higher-fidelity counterpart. This is reasonable in practice, since lower-fidelity simulations are by nature cheaper than higher-fidelity ones. The second condition is that the sample size $n_t$ at any node $t$ should be a multiple of the sample size $n_T$ at the root (highest-fidelity) node. This allows us to evenly allocate slices of design points over each node, thus maximizing the value from the sliced LHD structure. As we show later, this can be satisfied by simply rounding off the sample sizes optimized in the following subsection. 

\subsection{Sample size allocation given fixed computational budget} 

Consider next step (ii), the allocation of sample sizes $n_t$ given a fixed budget $C$. Let $C_t$ be the cost of performing a \textit{single} run at simulation node $t \in V$, and let $n_t$ be the allocated sample size at node $t$. The budget constraint can thus be written as $\sum_{t \in V} C_t n_t \leq C$. 

We now derive a design criterion to minimize under this constraint for the sample sizes $n_1, \cdots, n_T$. This follows from the proposition below (adapted from \citealp{wu1993local}), which provides an upper bound on prediction error from the GMGP model. In what follows, we denote the response surface at source nodes by $\delta_t(\bs{x})$, $t\in V_S$.

\begin{prop}
Suppose the highest-fidelity simulation $Z_T(\bs{x})$ follows the recursive model \eqref{eqn:GMGPR_Linear_rec}. Further suppose that: (i) the parameter space $\Omega$ is bounded and convex, and (ii) the discrepancy term $\delta_t(\bs{x})$ is in the native space $N_{r_t}(\Omega)$ equipped with norm $||\cdot||_{N_{r_t}(\Omega)}$, where $r_t$ (the correlation function in \eqref{eq:GMGPR_disc}) is the Mat\'ern kernel with smoothness parameter $\nu$. Then, using the posterior mean $m_{Z_T}(\bs{x})$ in \eqref{eqn:rec_pred}, the prediction error can be upper bounded by:
\begin{align}
\label{eq:errbd}
    |Z_T(\bs{x})-m_{Z_T}(\bs{x})| &\leq \sum_{t=1}^T\left(\prod_{t'\in \mathrm{Des}(t)\cup\{t\}\setminus\{T\}}|\rho_{t'}(\bs{{x})}|\right)c_{r_t}h_{\mathcal{D}_t}^{\nu}||\delta_t(\bs{x})||_{N_{r_t}(\Omega)}.
\end{align}
Here, $h_{\mathcal{D}_t} = \max_{\bs{x} \in \Omega} \min_{i=1, \cdots, n_t}\|\bs{x}-\bs{x}_i^t\|_2$ is the fill distance of design $\mathcal{D}_t$, and $c_{r_t}$ is a constant depending on the length-scale parameters of $r_t$.
\end{prop}
\noindent Our goal is to find an optimal allocation of sample sizes $n_1, \cdots, n_T$ to minimize the error bound on the right side of \eqref{eq:errbd}, which in turn reduces the GMGP prediction error.

However, prior to data, we do not know what $\rho_t(\bs{x})$ and $||\delta_t(\bs{x})||_{N_{r_t}(\Omega)}$ are, and thus require further assumptions to evaluate the desired error bound. Suppose we make the simplifying assumptions that, \textit{prior} to data, the dependency functions are constant over all nodes, i.e., $|\rho_t(\bs{x})| = \rho$, and that the native norm of the discrepancies are constant, i.e., $||\delta_t(\bs{x})||_{N_{r_t}(\Omega)} = \Delta$. Then, ignoring proportionality constants, the bound in \eqref{eq:errbd} reduces to $\sum_{t=1}^T \rho^{|\text{Des}(t)|} h_{\mathcal{D}_t}^{\nu}$. Additionally, if the design points $\mathcal{D}_t$ at \textit{each} node $t \in V$ are distributed in a manner which minimizes the fill distance asymptotically (this property is known as \textit{low-dispersion}; see, e.g., \citealp{fang1993number}), then it is known that $h_{\mathcal{D}_t} = \mathcal{O}(n_t^{-1/d})$ \citep{wendland2004scattered}. This low-dispersion property is satisfied by most space-filling criteria \citep{mMMm2017}. With this, the bound further simplifies to:
\begin{equation}
\sum_{t=1}^T \rho^{|\text{Des}(t)|} {n_t}^{-\nu/d} =: \Phi_\rho(n_1, \cdots, n_T),
\end{equation}
where $|\text{Des}(t)|$ is the number of descendant nodes on $t$. Consider now the minimization of the simplified bound $\Phi_\rho(n_1, \cdots, n_T)$ given the budget constraint, i.e.:
\begin{equation}
\min_{n_1, \cdots, n_T} \Phi_\rho(n_1, \cdots, n_T) \quad \text{s.t.} \quad \sum_{t \in V} C_t n_t \leq C.
\label{eq:sampleopt}
\end{equation}
Using the method of Lagrange multipliers with Karush-Kuhn-Tucker conditions \citep{nocedal2006numerical}, the optimal sample sizes can then be solved in closed-form as:
\begin{equation}\label{prop_design}
    n_t \propto \left(\frac{\rho^{|\text{Des}(t)|}}{C_t}\right)^{d/(d+\nu)}, \quad \sum_{t \in V} C_t n_t = C.
\end{equation}

A related Lagrange multiplier approach was employed for sample size allocation in multi-level Monte Carlo \citep{giles2008multilevel} and multi-level approximation \citep{sung2022stacking,ehara2023adaptive}, which are distinct from the current graphical problem. Equation \eqref{prop_design} provides a nice closed-form expression for sample size allocation over the multi-fidelity graph $\mathcal{G}$, where each node $t$ has a different cost $C_t$ per run. This yields two useful interpretations. First, nodes with higher costs per run are assigned smaller sample sizes by \eqref{prop_design}, which is intuitive since such experiments demand more computational resources. Second, assuming $\rho \in (0,1)$, nodes with a greater number of descendant nodes (i.e., those with more higher-fidelity refinements) are assigned smaller sample sizes. This is also intuitive: such nodes are correlated and share information with its many descendant nodes, and thus require fewer samples compared to a node with few descendants.

To evaluate the sample sizes in \eqref{prop_design}, however, we would need a prior estimate of the dependency parameter $\rho$. This may be obtained via a careful discussion with scientific modelers, to elicit the degree of dependency expected between simulation models prior to data. In the absence of such prior information, we suggest using a choice of $\rho$ between 0.5 and 0.9, which allows for some integration of the underlying graphical structure for experimental design. The optimized sample sizes from \eqref{prop_design} can then be used within the nested BFS design (Algorithm \ref{alg:DAGE}) to generate design points for multi-fidelity simulation. We show in Section \ref{sec:simulations} how this nested BFS design procedure with optimal sample size allocation can yield improved predictive performance for GMGP modeling.

\section{Numerical Experiments}
\label{sec:simulations}

We investigate the proposed GMGP models (r-GMGP and d-GMGP) compared to existing multi-fidelity models in a suite of numerical experiments. \cbl{Three metrics are used here:
\begin{itemize}
\itemsep0em 
    \item Root-mean-squared-error (RMSE): $\sqrt{M^{-1}\sum_{i=1}^{M}(\hat{y}_i-y_i)^2}$, where $y_i$ is the high-fidelity output for test point $i$ and $\hat{y}_i$ its prediction. This measures point prediction accuracy.
    \item Normalized root-mean-squared-error (N-RMSE): $1-{\text{RMSE}}/{\text{RMSE}_{\text{base}}}$, where $\text{RMSE}_{\text{base}}$ is the RMSE of the baseline sample mean predictor. This assesses point prediction accuracy relative to a baseline, with larger values indicating better predictions.
    \item Continuous Ranked Probability Score (CRPS; \citealp{gneiting2007strictly}): This measures the quality of \textit{probabilistic} predictions, with smaller values suggesting better predictions. Its specific expression can be found in \cite{gneiting2007strictly}.
\end{itemize}}

\noindent We first explore in Sections \ref{subsec:1dexp} and \ref{subsec:20dexp} the performance of the GMGP models with existing methods on a 1-d and 20-d synthetic function. We then investigate in Section \ref{subsec:expdesign} the performance of the proposed experimental design given varying budget constraints.


\subsection{1-dimensional experiment}
\label{subsec:1dexp}

\begin{figure}
\centering
\begin{minipage}{0.4\textwidth}
    \centering
    \begin{tikzpicture}
        \tikzset{vertex/.style = {shape=circle,draw,minimum size=1.5em}}
        \tikzset{edge/.style = {->,> = latex'}}
        \node[vertex] (a) at  (0,0) {$L_1$};
        \node[vertex] (b) at  (3,0) {$L_2$};
        \node[vertex] (c) at  (1.5,-1.5) {$H$};
        \draw[edge] (a) to (c);
        \draw[edge] (b) to (c);
    \end{tikzpicture}
\end{minipage}
\begin{minipage}{0.4\textwidth}
    \centering
    \includegraphics[width=\linewidth]{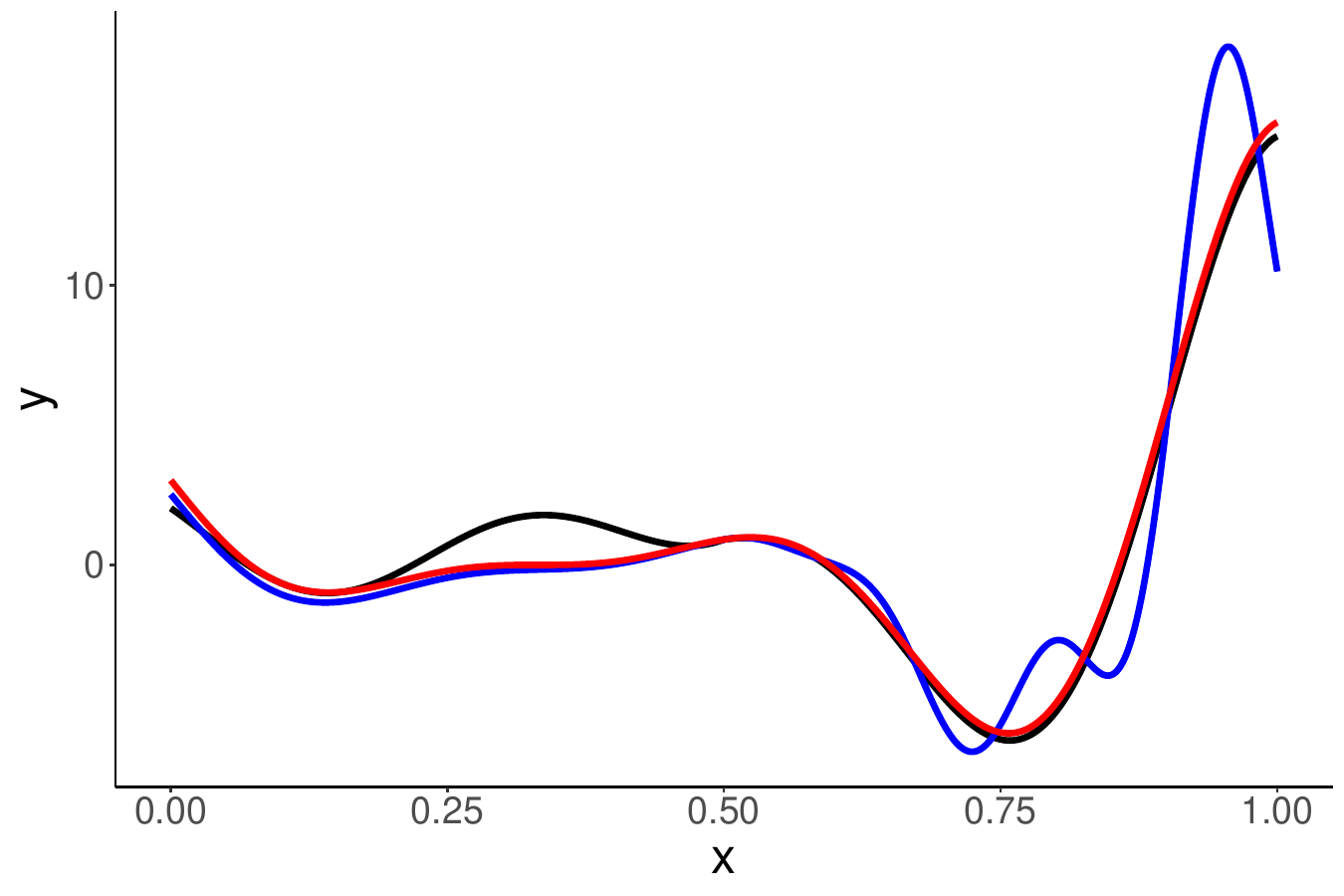}
\end{minipage}
\caption{Visualizing the 1-d experiment: \textbf{(left)} the 3-node DAG used for generating the true functions; \textbf{(right)} the true high-fidelity (red),  low-fidelity 1 (blue), and low-fidelity 2 (black) functions.}
\label{fig:1dexperiment}
\end{figure}

We first consider a simulation study using the simple 3-node graph in Fig. \ref{fig:1dexperiment} (left), with three correlated $d=1$-dimensional functions. The high-fidelity function (denoted $H$) is taken to be the Forrester function \citep{Forrester2008} over the design space $[0,1]$:
\[Z_H(x) = (6x-2)^2 \sin(12x-4),\]
and the two low-fidelity functions ($L_1$ and $L_2$) are derived from $H$ as follows:
\small
\begin{equation*}
    \begin{cases}
        Z_{L_1}(x) = \mathbbm{1}_{\{x<0.5\}}\left[Z_H(x) + (x-0.5)\right] + \mathbbm{1}_{\{x\geq 0.5\}}\left[Z_H(x) + (x-0.5) \cos(40x) (5x-1)^2\right] \\
        Z_{L_2}(x) = \mathbbm{1}_{\{x\leq0.5\}} \left[Z_H(x) + 2(x-0.5)\cos(10x)
        (10x-1)^2 \right] + \mathbbm{1}_{\{x>0.5\}} \left[Z_H(x) - (x-0.5)\right].
    \end{cases}
\end{equation*}
\normalsize
\noindent Fig. \ref{fig:1dexperiment} (right) visualizes the three test functions. Here, $L_1$ is designed to be closer to $H$ on $[0,0.5)$, while $L_2$ is closer to $H$ on $(0.5,1]$. This mimics a scenario where, due to simplifications in the underlying simulation model, low-fidelity functions might capture well the high-fidelity function in certain regions of the parameter space but not within other regions. \cbl{Here, the high-fidelity function is slightly smoother than the two low-fidelity functions, which may arise when the lower-fidelity simulator introduces spurious high frequencies from grid discretization \citep{okuda1972nonphysical}.} The training sample sizes for $L_1$, $L_2$ and $H$ are set as 15, 15 and 8, respectively, with the design points in $H$ nested within $L_1$ and $L_2$. For testing, $M=1000$ evenly-spaced points on $[0,1]$ are used. 

We compare 6 different predictive models here: the ``high-fidelity'' GP, two variants of the KO model, the NARGP (nonlinear autoregressive multi-fidelity GP regression) model \citep{Perdikaris_2017}, and the proposed r-GMGP and d-GMGP models. For the high-fidelity GP, only the 8 points on $H$ are used for model training. Recall that a key limitation of the KO model is that the simulations (see Fig. \ref{fig:1dexperiment} (left)) cannot be ranked from lowest to highest fidelity. To that end, two variants of the KO model are considered. The first is the KO-path model (as discussed in Section \ref{sec:ko}), where the KO model is fit using only data along the longest path (here, $L_1 \rightarrow H$). The second is the KO-misspecified model, where the KO model is fit on the \textit{full} simulation data with an arbitrary ordering from lowest to highest fidelity (here, $L_2 \rightarrow L_1 \rightarrow H$). Since the NARGP model also requires such a ranking, this model is fit on data from $L_1$ and $H$ (similar to KO-path). Both the proposed r-GMGP and d-GMGP models make use of the full training dataset along with the implicit DAG structure for simulation. We apply the kernel in \cite{Perdikaris_2017} for NARGP, the kernel in \eqref{eqn:d-GMGP_kernel} for d-GMGP, the squared-exponential kernel for the high-fidelity GP\footnote{\cbl{The fit with a Mat\'ern kernel here yielded poor results, hence our use of the squared-exponential kernel.}}, and the Mat\'ern-5/2 kernel for the remaining models.

\begin{figure}[!t]
    \centering
    \includegraphics[width=.99\linewidth]{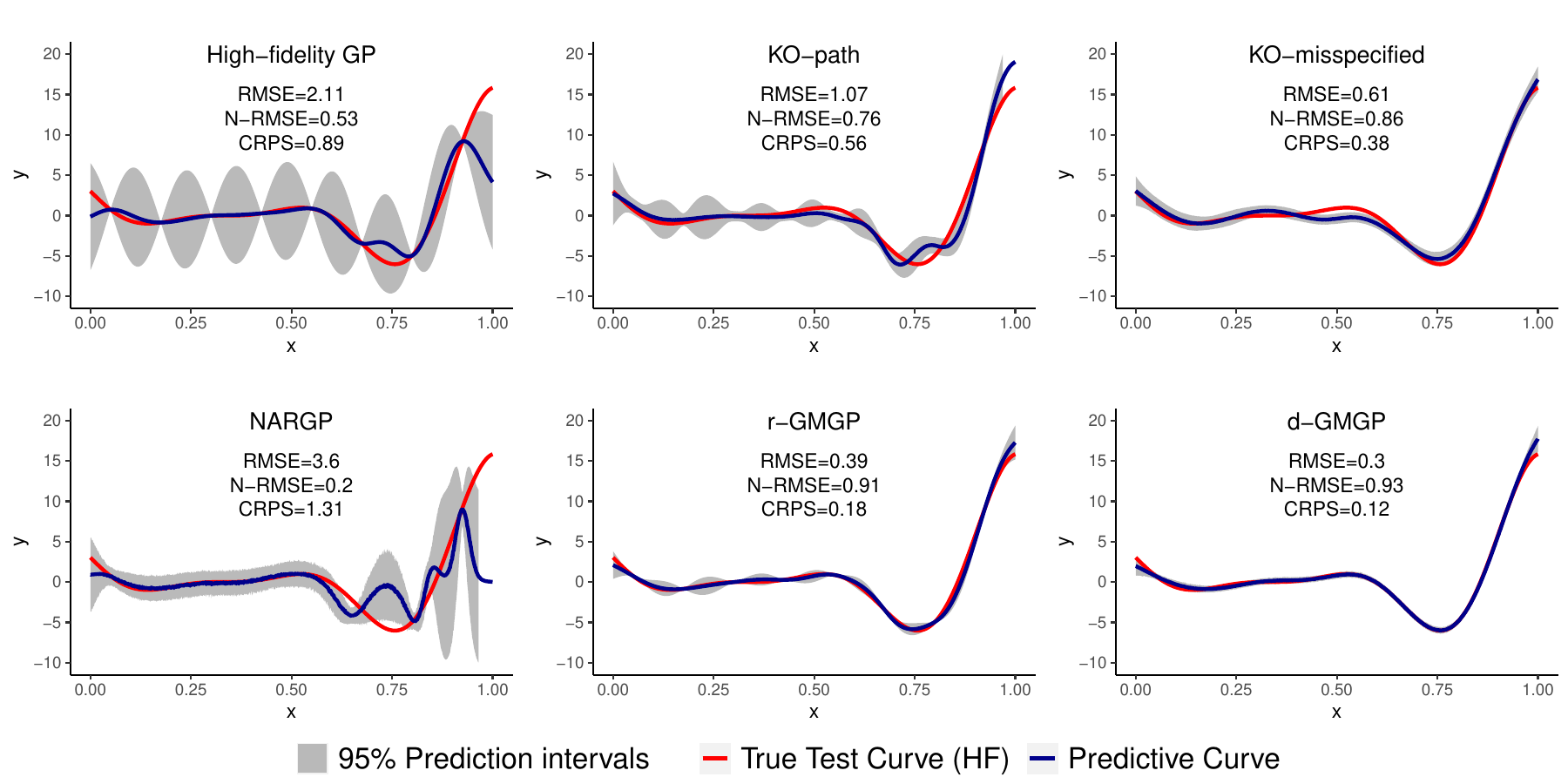}
  \caption{Results for the 1-d experiment: predictions (dark blue), 95\% predictive intervals (gray) and the true high-fidelity function (red), along with its predictive metrics.}
  \label{fig:Example1d}
\end{figure}



Fig. \ref{fig:Example1d} compares the predictive performance of the aforementioned predictive models. We see that the proposed r-GMGP and d-GMGP models yield noticeably improved predictions over existing models, in terms of both metrics. This improvement suggests that, when the underlying graphical model dependency structure is known from prior scientific knowledge, incorporating such structure can lead to better predictive performance. Comparing the two GMGP models, we see that d-GMGP slightly outperforms r-GMGP. This is not too surprising, since from Fig. \ref{fig:1dexperiment} (right), we see that the dependency between nodes is quite nonlinear between the low and high fidelity functions, so a more flexible (nonlinear) structure should yield improved predictive models. The online supplement reports similar results for a 5-d experiment using the same DAG.

\subsection{20-dimensional experiment}
\label{subsec:20dexp}


We then investigate performance on a higher-dimensional problem with highly nonlinear dependencies between nodes. We use here the 5-node DAG from Fig. \ref{fig:motivation} (left), which consists of two low-fidelity functions ($L_1$ and $L_2$), two medium-fidelity functions ($M_1$ and $M_2$) and one high-fidelity function ($H$). The high-fidelity function $H$ is taken to be the $d=20$-dimensional test function from \cite{WELCH1992} over design space $[-0.5,0.5]^{20}$. The medium-fidelity function $M_1$ is obtained by \textit{averaging} $H$ over a sliding window of width $\pm 0.1x_l$ over each input $l=1, \cdots, 20$. The low-fidelity functions $L_1$ and $L_2$ are similarly obtained by averaging $M_1$ over a sliding window of size $\pm 0.15x_l$, over the 10 \textit{odd} inputs (i.e., $x_1, \cdots, x_{19}$) for $L_1$, and over the 10 \textit{even} inputs for $L_2$. This mimics the scenario where lower-fidelity functions are obtained via an averaging operation, which is widely encountered in physics (e.g., \citealp{pope2000turbulent}). The remaining medium-fidelity function $M_2$ is obtained via a simple approximation of $H$: $Z_{M_2}(\bs{x}) = 1.2 Z_H(\bs{x})-1$. Design points for $L_1$, $L_2$, $M_1$, $M_2$ and $H$ are generated from a maximin SLHD with sample sizes 200, 200, 160, 160, and $n_H$, respectively, where the high-fidelity sample size $n_H$ increases from 40 to 120 in increments of 20. For testing, $M=500$ random test samples are used. This procedure is repeated for 20 times to account for error variability. Since dependencies between nodes are highly nonlinear, the d-GMGP is used in place of the r-GMGP model.

\begin{figure}[!t]
    \centering
    \includegraphics[width=.9\linewidth]{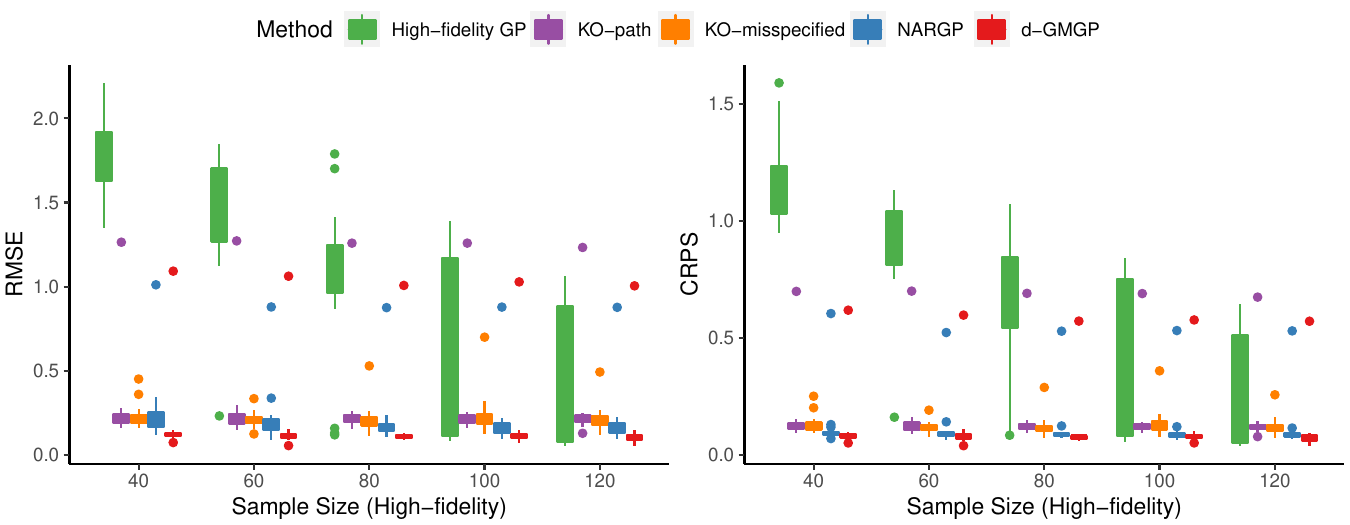}
  \caption{Results for the 20-d experiment: boxplots of two performance metrics for different sample sizes on $H$ (\textbf{left}: RMSE, \textbf{right}: CRPS).}
  \label{fig:PerformanceComparison}
\end{figure}

Fig. \ref{fig:PerformanceComparison} shows the boxplots of \cbl{RMSE and CRPS (N-RMSE results are in the online supplement)} for various high-fidelity sample sizes. There are several observations of interest. First, for all sample sizes, d-GMGP outperforms existing models, which again shows that, by leveraging the underlying model dependency structure in the form of a DAG, the proposed method can yield significant improvements in terms of predictive performance. Second, this improvement is most pronounced when the high-fidelity sample size is small. This is intuitive: as high-fidelity data become limited, additional structure linking multi-fidelity data should be more effective in improving predictive performance. Here, the d-GMGP model seems capable of leveraging this DAG dependency structure for prediction, yielding good performance even in the challenging setting where there is limited high-fidelity training data in a high-dimensional space.

\subsection{Experimental design}
\label{subsec:expdesign}

Finally, we investigate the proposed designs from Section \ref{sec:Design} on the earlier 1-d and 5-d problems (the latter in online supplement), which use the 3-node DAG in Fig. \ref{fig:1dexperiment} (left). For the 1-d problem, we set the computational cost per run at each node ($L_1$,$L_2$,$H$) to be $(2,2,32)$; for the 5-d problem, this cost per run is set as $(2,2,64)$, respectively. We then compare the design methods (discussed below) on cost budgets ranging from 160 to 360 in increments of 40 for the 1-d problem, and from 600 to 1200 in increments of 100 for the 5-d problem.

The proposed design from Section \ref{sec:Design} is compared to two baseline design approaches given a fixed computational budget $C$. The first approach allocates the \textit{full} budget to the high-fidelity node $H$. A Sobol' sequence \citep{joe2003remark} is used for the 1-d problem, and a maximin LHD \citep{morris1995exploratory} is used for the 5-d problem. The second approach allocates the computational budget to the three nodes $L_1$, $L_2$ and $H$ with a \textit{fixed} ratio of sample sizes. Two choices of fixed ratios are used: $6:6:1$ and $8:8:1$ for the 1-d problem, and $2:2:1$ and $3:3:1$ for the 5-d problem. The ratios are chosen such that the budget allocation is likely to be different from the proposed design. For the 1-d problem, a Sobol' sequence is used to generate the nested designs over the DAG (see Section \ref{sec:bfs} for details); for the 5-d problem, a maximin SLHD \citep{Ba2015SLHD} is used to generate the nested designs. We then implement the proposed design approach with moderate dependency ($\rho=0.5$ for 1-d, $\rho=0.6$ for 5-d) and high dependency ($\rho=0.9$), by first computing the desired sample size at each node via \eqref{prop_design}, then allocating the design points using the nested BFS design (Algorithm \ref{alg:DAGE}). The standard GP model and r-GMGP model with Mat\'ern 5/2 kernel are then fit to the data, and the performances are compared via the RMSE of $M=100$ test points for 1-d problem and $M=500$ test points for 5-d problem. We repeat the procedure 20 times to reduce sampling variation.

\begin{figure}[!t]
    \centering
    \includegraphics[width=0.85\linewidth]{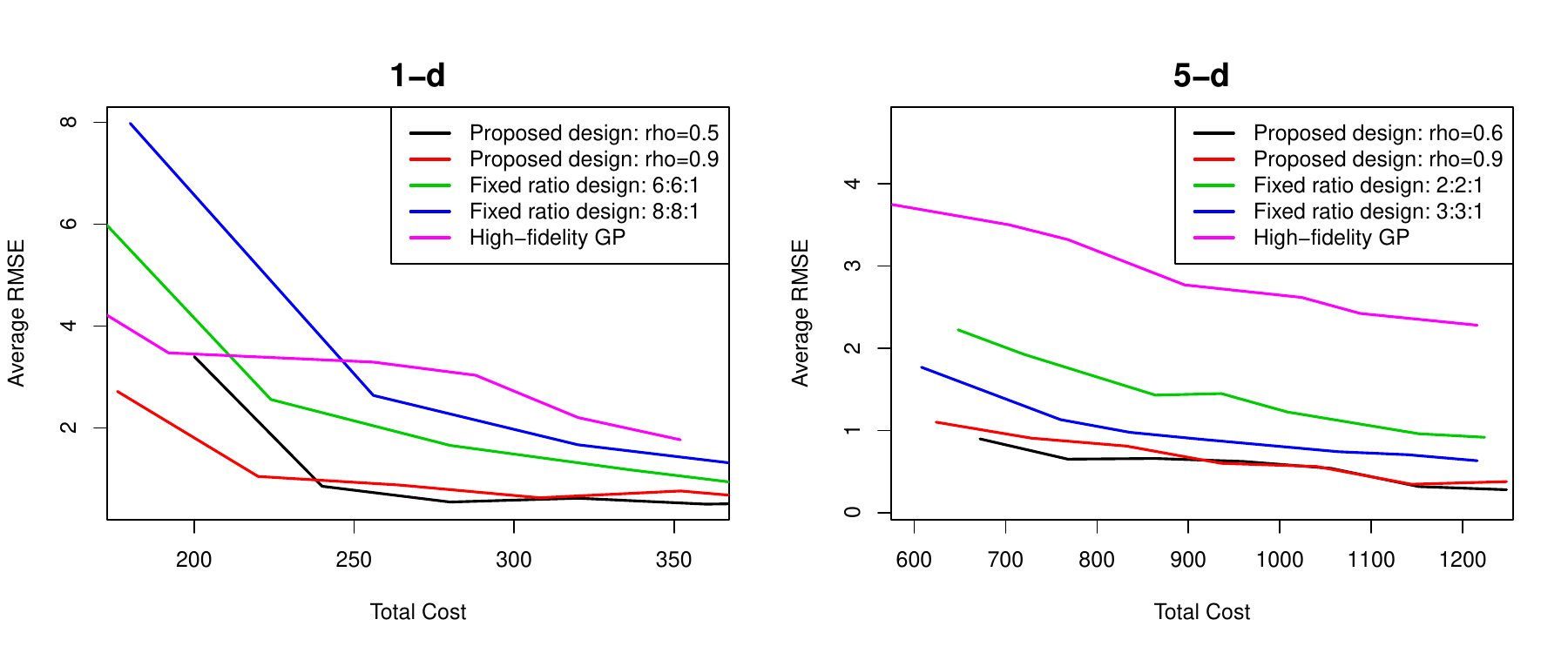}
    \vspace{-0.5cm}
  \caption{A comparison of average RMSE for five considered design procedures, for the 1-d problem (\textbf{left}) and the 5-d problem (\textbf{right}).}
  \label{fig:Example5dDesigns}
\end{figure}

Fig. \ref{fig:Example5dDesigns} shows the average RMSE of the compared design approaches for the 1-d and 5-d problems. We see that, given a fixed budget $C$, the proposed designs for both moderate and high dependencies yield noticeably better predictive performance to both the high-fidelity design (where the full budget is allocated to high-fidelity runs) and the fixed ratio designs. This suggests that the proposed design approach, which jointly determines sample sizes and allocates sample points over each node in the DAG, is quite effective in reducing predictive error given a fixed budget, which is as desired. Furthermore, from these (and other) experiments, the performance of our designs appear quite robust to the choice of dependency parameter $\rho$, given it is sufficiently large (but not equal to 1). In applications where one expects a reasonable degree of dependency between simulations a priori, we would expect the proposed design with $\rho \in [0.5,0.9]$ to yield better performance over a design with arbitrarily fixed ratio, and certainly over a design with only high-fidelity samples.

\section{Multi-fidelity emulation of heavy-ion collisions} \label{sec:app}

We now return to the motivating problem for emulation of heavy-ion collisions. Experiments at Brookhaven National Laboratory and the European Organization for Nuclear Research (CERN) study collisions of atomic nuclei at velocities close to the speed of light. The temperature and pressure that the colliding nuclei are subjected to in these collisions converts them into a plasma of subatomic particles. The study of nuclear collisions has increasingly employed the use of \textit{computer} simulations, which involve complex scientific models that describe successively different stages of the collision. Such computer experiments are then integrated with the limited physical experimental data for calibration of unknown physical parameters. In recent years, Bayesian inference with such multi-stage simulations have led to novel discoveries on nuclear plasmas \citep{Bernhard:2019bmu,everett2021phenomenological,everett2021multisystem}. The computational cost of these simulations is considerable, however, requiring thousands of CPU hours per run over a high-dimensional input space. This presents a crucial bottleneck for a full-scale study of nuclear collisions, and emulator models are widely used to address this computational constraint~\citep{Novak:2013bqa}.


The computer simulators for heavy-ion collisions naturally form a multi-stage model~\citep{everett2021multisystem} as shown in Fig.~\ref{fig:intro_flowchart}. In our study, we consider three broad stages. The first stage models the initial impact of the two nuclei prior to hydrodynamic evolution. This pre-hydrodynamic phase can be simulated by the computationally efficient Trento model~\citep{Moreland:2014oya}. The second stage involves the hydrodynamic evolution of the quark-gluon-plasma. A full-fledged simulation requires numerically expensive 3D hydrodynamic modeling (which is too expensive for our study), so we neglect viscosity and consider ``3+1D \emph{ideal} QCD hydrodynamics'' as the highest-fidelity model in this stage. This stage can be simplified by reducing the dimensionality of the hydrodynamic simulation from 3D to 1D, resulting in a lower-fidelity ``1+1D ideal QCD hydrodynamics" model. A further simplification is the ``1+1D linearized ideal conformal hydrodynamics" model, which replaces the equation of state of nuclear matter by a simpler conformal equation of state, and employs a linearization of the hydrodynamics equations. The third stage models the post-hydrodynamic conversion of the nuclear fluid into particles, using the Cooper-Frye prescription~\citep{Cooper:1974mv}. For the observables of interest in this study, it is possible to omit this conversion in lower fidelity models. The multi-stage simulation framework is visualized in Fig. \ref{fig:intro_flowchart} and described in \cite{everett2021multisystem}. 

This multi-stage framework provides a rich testbed for multi-fidelity simulations, since experiments of different fidelities can be performed via different model combinations at each stage. After close discussions, we decided to run the following three models:
\begin{itemize}
\itemsep0em 
    \item $L_1$: Pre-hydrodynamics + linearized ideal conformal hydrodynamics + Cooper-Frye,
    \item $L_2$: Pre-hydrodynamics + 1+1D ideal QCD hydrodynamics,
    \item $H$: Pre-hydrodynamics + 3+1D ideal QCD hydrodynamics + Cooper-Frye.
\end{itemize}
Here, $H$ is the highest-fidelity model, and $L_1$ and $L_2$ are lower-fidelity representations of $H$. The corresponding DAG for this experiment is given in Fig. \ref{fig:1dexperiment} (left).

Table \ref{tab:QGP_APP} (left) summarizes the $d=9$ input parameters in this computer experiment, which are shared by all three simulation models. Here, we study a single output (or ``observable''), which is the ratio of pions produced at rapidity $y=0$ and rapidity $y=1$. To train the emulator, we first simulate 25, 200 and 200 training points from $H$, $L_1$ and $L_2$, respectively. These design points are obtained via a maximin SLHD design \citep{Ba2015SLHD}, such that the design for $H$ is nested within that for $L_1$ and $L_2$ (see Section \ref{sec:bfs} for details). To evaluate predictive performance, $M=75$ out-of-sample testing points are generated from a separate maximin LHD design. As before, we compare the proposed r-GMGP and d-GMGP with the high-fidelity GP, the two KO model variants (KO-path and KO-misspecified), and the NARGP model in \cite{Perdikaris_2017}. \cbl{Kernel choices for each model are the same as in earlier numerical experiments (Section \ref{sec:simulations}). All models are fitted in {Python}, except for the r-GMGP which uses \textsc{R}.}

\begin{table}
\parbox{.48\textwidth}{
\centering
\begin{adjustbox}{max width=0.47\textwidth}
\begin{tabular}{p{6cm} c}
    \toprule
    {Description} & {Range} \\
    \midrule
     {nucleon width} & {[0.35,1.40](fm)} \\
    {constituent width fraction} & {[0.1,0.9]} \\
    {min transverse kinematic cut} & {[0.2,1.0](GeV)} \\
     {fragment profile shape $\alpha$} & {[3.0,5.0]}  \\
    {fragment profile shape $\beta$} & {[-1.0,0.0]} \\
    {midrapidity energy density power} & {[0.30,0.48]} \\
    {midrapidity energy density normalization} & {[0.20,0.45]} \\
   {fireball--fragment fluctuation} & {[0.1,0.6]} \\
   {fireball profile shape} & {[1.0,1.5]} \\
    \bottomrule
\end{tabular}
\end{adjustbox}
}
\hspace{0.3cm}
\parbox{.48\textwidth}{
\centering
\begin{adjustbox}{max width=0.47\textwidth}
\begin{tabular}{l c c c c}
    \toprule
    \multicolumn{1}{c}{} & 
    \multicolumn{3}{c}{Metrics ($\times 10^{-2}$, except N-RMSE)} & 
    \multicolumn{1}{c}{Computation}\\
    \cmidrule(lr){2-5}  
    \multicolumn{1}{c}{Model} & {RMSE} & {N-RMSE} & {CRPS} & {Time (s)} \\
    \midrule
    {High-fidelity GP} & {5.49} & {0.46} & {3.54} & {2.83}\\
    {KO-path} & {3.48} & {0.66} & {1.99} & {5.97}\\
    {KO-misspecified} & {3.95} & {0.71} & {2.30} & {10.13} \\
    {NARGP} & {3.66} & {0.73} & {2.13} & {92.15} \\
    {r-GMGP} & {2.92} & {0.72} & {1.64} & {3.08} \\
    {d-GMGP} & {2.17} & {0.79} & {1.34} & {135.69} \\
    \bottomrule
\end{tabular}
\end{adjustbox}
}
\caption{(\textbf{left}) $d=9$ input parameters for the heavy-ion collision simulator. (\textbf{right}) Predictive performance metrics and computational times for the heavy-ion collision application.}
\label{tab:QGP_APP}
\end{table}

Table \ref{tab:QGP_APP} (right) reports the predictive performance metrics and computational time for the considered emulator models. We see that the computational times for NARGP and d-GMGP models are longer than other methods, which is unsurprising since both approaches require Monte Carlo approximation. Among all six models, the proposed r-GMGP and d-GMGP give better predictive performance than existing methods, providing noticeable reductions over both test error metrics. This again suggests that, by integrating the DAG dependency structure between the simulators (i.e., the ``science'') as prior knowledge, one can greatly improve predictive performance. This ``science'' appears to play a key role: compared to the KO-misspecified model, which misspecifies the underlying scientific connections between simulation models, the proposed GMGP models yield noticeable improvements in predictive performance by integrating model dependency information elicited from a careful inspection of the modeled physics. This can then be used for improved parameter constraints on nuclear plasma properties, leading to more precise scientific discoveries given a fixed experimental budget.

\section{Conclusion}
\label{sec:concl}

We proposed a new Graphical Multi-fidelity Gaussian Process (GMGP) model for multi-fidelity predictive modeling. The key novelty of the GMGP is the integration of scientific information in the form of a DAG, which captures connections between different simulation models in terms of fidelities. This DAG should be obtained via a careful inspection of the scientific simulation models and discussion with domain scientists. We show that the GMGP model has appealing properties for multi-fidelity modeling, and present two extensions which allow for nonlinear modeling and scalable prediction via recursive computation of the posterior predictive mean and variance \cbl{at each depth level of the DAG}. We also present a comprehensive experimental design methodology for the GMGP, which jointly determines the sample size on each simulation model and its corresponding design points over the parameter space. Extensive numerical experiments and an application in heavy-ion collisions demonstrate the improvement of the proposed method over existing multi-fidelity models, particularly given a tight experimental budget for simulations.

There are several interesting avenues for future work. One current limitation is that the equivalence of GMGP and r-GMGP is only shown for directed in-trees. Although this can often be satisfied via a careful design of the lower-fidelity models, it may be violated when the multi-fidelity training data are \textit{observed} rather than designed. It would thus be of interest to extend this recursive formulation for more general graphs, \cbl{and the incorporation of additional DAG structure under nested designs seems promising}. \cbl{For problems where the multi-fidelity DAG is not known with certainty, it would be useful to explore the use of DAG learning algorithms (e.g., \citealp{you2018graphrnn,zhang2019d}) within the GMGP.} Another intriguing direction would be to explore an efficient, fully Bayesian implementation that accounts for uncertainties from all model parameters; the recent work of \cite{ma2020objective} seems to be promising for this direction. \cbl{The exploration of more flexible sample sizes at different nodes is also of interest, and recent developments in flexible sliced designs \citep{kong2018flexible} appear fruitful. Finally, in applications where model observables are highly non-stationary and/or discontinuous, the extension of the GMGP using binary tree partitions \citep{konomi2021bayesian} would be useful.} 



\noindent \textbf{Supplementary Materials:} The online supplementary materials include proofs for Propositions 1 and 2, the algorithm for d-GMGP prediction, two additional figures for the numerical experiments in Section \ref{sec:simulations}, and code for reproducing Figure \ref{fig:Example1d}.

\if0\blind{
\noindent \textbf{Acknowledgements}: The authors gratefully acknowledge funding from NSF CSSI 2004571, NSF DMS 2210729, NSF DMS 2316012, DE-SC0024477 (YJ, SM), and DE-FG02-05ER41367 (SAB, JFP and DS). We also thank the JETSCAPE collaboration (\url{https://jetscape.org/}) for insightful conversations and discussions.
}
\fi

\newpage
\spacingset{1.0}
\bibliography{references}

\section{Supplementary Material}
The supplementary material contains proofs for Propositions 1 and 2, the algorithm for d-GMGP prediction and two additional figures for the numerical experiments in Section 5.

\section{Proof of Proposition 1}

\subsection{Property (a)}
Property (a) can be derived from the well-known Factorization Theorem for Bayesian networks \citep{russell03a}. Let $\{Z_1(\bs{x}),Z_2(\bs{x}),\cdots,Z_T(\bs{x})\}$ denote the simulation outputs with input $\bs{x}$ at all nodes on a DAG, where $Z_t(\bs{x})$ is the output at node $t$. By assuming a Bayesian Network structure over the DAG, the joint distribution of observations can be factorized as
\begin{align*}
    P(Z_1(\bs{x}),Z_2(\bs{x}),\cdots,Z_T(\bs{x})) = \prod_{i=1}^{T}P(Z_i(\bs{x})|\{Z_j(\bs{x})\}_{j\in \text{Pa}(i)}).
\end{align*}

\noindent Then, for all $Z_t(\bs{x})$, we have:
\begin{align*}
    P(Z_t(\bs{x})|\{Z_j(\bs{x})\}_{j\in \text{Anc}(t)}) &= \frac{P(Z_t(\bs{x}),\{Z_j(\bs{x})\}_{j\in \text{Anc}(t)})}{P(\{Z_j(\bs{x})\}_{j\in \text{Anc}(t)})}\\
    &= \frac{\prod_{l\in \text{Anc}(t)\cup\{t\}}P(Z_l(\bs{x})|\{Z_m(\bs{x})\}_{m\in \text{Pa}(l)})}{\prod_{l\in \text{Anc}(t)}P(Z_l(\bs{x})|\{Z_m(\bs{x})\}_{m\in \text{Pa}(l)})}\\
    &= P(Z_t(\bs{x})|\{Z_j(\bs{x})\}_{j\in \text{Pa}(t)}).
\end{align*}

\noindent Thus, conditioning on parent nodes of $t$, it follows that $Z_t(\bs{x})$ and ancestors of $t$ that are not parent nodes are independent, i.e., $Z_t(\bs{x}) \perp Z_{t'}(\bs{x}) | \{Z_j(\bs{x})\}_{j \in \textup{Pa}(t)}$ for all $t'\neq t, t' \notin \textup{Des}(t), t'\notin \textup{Pa}(t)$.

\subsection{Property (b)}
Next, we present the proof of Property (b) for the simplest 3-node in-tree (Fig. \ref{fig:APP_simpleGraph}). This approach can be extended analogously for more complicated in-trees. 
\begin{figure}[H]
    \centering
    \begin{tikzpicture}
        \tikzset{vertex/.style = {shape=circle,draw,minimum size=1.5em}}
        \tikzset{edge/.style = {->,> = latex'}}
        \node[vertex] (a) at  (0,0) {1};
        \node[vertex] (b) at  (2,0) {2};
        \node[vertex] (c) at  (1,-1) {3};
        \draw[edge] (a) to (c);
        \draw[edge] (b) to (c);
    \end{tikzpicture}
    \caption{Simplest 3-node in-tree.}\label{fig:APP_simpleGraph}
\end{figure}

\textit{Proof.} The GMGP model for a 3-node in-tree is given as follows:
\begin{align*}
    \begin{cases}
        Z_3(\bs{x}) = \rho_1(\bs{x})\cdot Z_1(\bs{x}) + \rho_2(\bs{x})\cdot Z_2(\bs{x}) + \delta_3(\bs{x})\\
        \delta_3(\bs{x})\perp Z_1(\bs{x}),\ \delta_3(\bs{x})\perp Z_2(\bs{x}),\ Z_1(\bs{x})\perp Z_2(\bs{x}).
    \end{cases}
\end{align*}

\noindent Thus we have for $\bs{x}'\neq \bs{x}$,
\begin{align*}
    \text{Cov}(Z_3(\bs{x}),Z_1(\bs{x}')) = \text{Cov}(\rho_1(\bs{x})\cdot Z_1(\bs{x}), Z_1(\bs{x}'))
    = \rho_1(\bs{x})\cdot \text{Cov}(Z_1(\bs{x}),Z_1(\bs{x}')),
\end{align*}
\noindent and
\small
\begin{align*}
    &\quad \left[\text{Cov}(Z_3(\bs{x}),Z_1(\bs{x})),\text{Cov}(Z_3(\bs{x}),Z_2(\bs{x}))\right]
    \begin{bmatrix}
        \text{Var}(Z_1(\bs{x}))^{-1} & 0\\
        0 & \text{Var}(Z_2(\bs{x}))^{-1}
    \end{bmatrix}
    \begin{bmatrix}
        \text{Cov}(Z_1(\bs{x}'),Z_1(\bs{x}))\\
        \text{Cov}(Z_1(\bs{x}'),Z_2(\bs{x}))
    \end{bmatrix}\\
    &= \text{Cov}(Z_3(\bs{x}),Z_1(\bs{x}))\text{Var}(Z_1(\bs{x}))^{-1}\text{Cov}(Z_1(\bs{x}),Z_1(\bs{x}')) + \text{Cov}(Z_3(\bs{x}),Z_2(\bs{x}))\text{Var}(Z_2(\bs{x}))^{-1}\text{Cov}(Z_2(\bs{x}),Z_1(\bs{x}'))\\
    &= \rho_1(\bs{x})\cdot \text{Var}(Z_1(\bs{x}))\text{Var}(Z_1(\bs{x}))^{-1}\text{Cov}(Z_1(\bs{x}),Z_1(\bs{x}')) + 0\\
    &= \rho_1(\bs{x})\cdot \text{Cov}(Z_1(\bs{x}),Z_1(\bs{x}')).
\end{align*}
\normalsize

\noindent From the above two expressions, we have
\begin{align*}
    &\quad \ \text{Cov}(Z_3(\bs{x}),Z_1(\bs{x}')|Z_1(\bs{x}),Z_2(\bs{x}))\\ &= \rho_1(\bs{x})\cdot \text{Cov}(Z_1(\bs{x}),Z_1(\bs{x}')) - \rho_1(\bs{x})\cdot \text{Cov}(Z_1(\bs{x}),Z_1(\bs{x}'))\\
    &= 0.
\end{align*}
\noindent Similarly, we can show that $\text{Cov}(Z_3(\bs{x}),Z_2(\bs{x}')|Z_1(\bs{x}),Z_2(\bs{x}))=0,\ \bs{x}'\neq\bs{x}$. This approach can be naturally extended for general in-trees, and thus Property (b) in Proposition 1 then follows. 

\newpage

\section{Proof of Proposition 2}

We adopt an inductive approach to prove Proposition 2. In Subsection \ref{Sec1.2.1} we first show that the equivalence between the GMGP and r-GMGP model for the simplest 3-node in-tree of depth 2 (see Fig. \ref{fig:APP_simpleGraph}), and argue that this naturally extends for in-trees of depth 2 with more than two branches (see Fig. \ref{fig:APP_extendedGraph}). This establishes the base case for induction. In Subsection \ref{Sec1.2.2}, we then show this equivalence for in-trees constructed by connecting two lower-level in-trees to a root node (see Fig. \ref{fig:APP_connectedGraph} (left)), and again argue that this extends for in-trees connected by more lower-level in-trees (see Fig. \ref{fig:APP_connectedGraph} (right)). This provides the inductive step. Finally, to complete the proof, we leverage the fact that any directed in-trees can be recursively constructed via the construction in Fig. \ref{fig:APP_connectedGraph} (right), thus proving the equivalence between the GMGP and r-GMGP for all directed in-trees. 

\subsection{Sub-graphs starting with source nodes}\label{Sec1.2.1}

\textit{Proof.} Let us again consider the simplest 3-node in-tree, as shown in Fig. \ref{fig:APP_simpleGraph}. We extend the proof in \cite{LeGratiet_2014} as follows. Note that the proof below naturally generalizes to the case of in-trees with multiple source nodes (see Fig. \ref{fig:APP_extendedGraph}).

\begin{figure}[H]
    \centering
    \begin{tikzpicture}
        \tikzset{vertex/.style = {shape=circle,draw,minimum size=1.5em}}
        \tikzset{edge/.style = {->,> = latex'}}
        \node[vertex] (a) at  (0,0) {1};
        \node[vertex] (b) at  (1,0) {2};
        \node[vertex] (c) at  (4,0) {\tiny{$T-1$}};
        \node[vertex] (d) at  (2,-1) {$T$};
        \draw[edge] (a) to (d);
        \draw[edge] (b) to (d);
        \draw[edge] (c) to (d);
        \draw[loosely dotted] (b) to (c);
    \end{tikzpicture}
    \caption{In-trees of depth 2 with multiple source nodes.}\label{fig:APP_extendedGraph}
\end{figure}

Let $\mathcal{D}_s$ be the design points at node $s,\ s=1,2,3$, and let $Z_s$ be the computer code at node $s$. Let $\bs{V}^{(s,s)}=\text{Var}(Z_s(\mathcal{D}_s))$ be the covariance matrix of $Z_s(\mathcal{D}_s)$, and $\bs{V}_s$ be the full covariance matrix of the data generated by the simulators from node 1 to node $s$. With the independence assumption between codes $Z_1$ and $Z_2$, we have 
\begin{align*}
\bs{V}_3 = 
    \begin{bmatrix}
        \bs{V}^{(1,1)} & \bs{0} & \bs{U}^{(1,3)}\\
        \bs{0} & \bs{V}^{(2,2)} & \bs{U}^{(2,3)}\\
        \bs{U}^{(1,3)^T} & \bs{U}^{(2,3)^T} & \bs{V}^{(3,3)}
    \end{bmatrix} =
    \begin{bmatrix}
        \bs{V}_2 & \bs{U}_2\\
        \bs{U}_2^T & \bs{V}^{(3,3)}
    \end{bmatrix},
\end{align*}
where $\mathcal{D}_1$ and $\mathcal{D}_2$ are ordered such that $\mathcal{D}_1=[\mathcal{D}_1\setminus \mathcal{D}_3,\mathcal{D}_3],\ \mathcal{D}_2=[\mathcal{D}_2\setminus \mathcal{D}_3,\mathcal{D}_3]$, and $\bs{U}_2=\left[\bs{U}^{(1,3)},\bs{U}^{(2,3)}\right]^T$. Here, $\bs{U}^{(1,3)}$ and $\bs{U}^{(2,3)}$ are covariances defined as:

\begin{align*}
\begin{cases}
    \bs{U}^{(1,3)} &= \text{Cov}(Z_1(\mathcal{D}_1),Z_3(\mathcal{D}_3)) 
    = (\bs{1}_{n_1}\bs{\rho}_1(\mathcal{D}_3)^T)\odot \bs{V}^{(1,1)}(\mathcal{D}_1,\mathcal{D}_3)\\
    \bs{U}^{(2,3)} &= (\bs{1}_{n_2}\bs{\rho}_2(\mathcal{D}_3)^T)\odot \bs{V}^{(2,2)}(\mathcal{D}_2,\mathcal{D}_3),
\end{cases}
\end{align*}
where $\bs{V}^{(1,1)}(\mathcal{D}_1,\mathcal{D}_3)$ and $\bs{V}^{(2,2)}(\mathcal{D}_2,\mathcal{D}_3)$ contain the last $n_3$ columns of $\bs{V}^{(1,1)}$ and $\bs{V}^{(2,2)}$, respectively.

Thus, $\bs{V}_2^{-1}\bs{U}_2$ can be simplified to
\begin{align*}
    \bs{V}_2^{-1}\bs{U}_2 &= \begin{bmatrix}
        (\bs{V}^{(1,1)})^{-1} & \bs{0}\\
        \bs{0} & (\bs{V}^{(2,2)})^{-1}
    \end{bmatrix}
    \begin{bmatrix}
        \bs{U}^{(1,3)}\\
        \bs{U}^{(2,3)}
    \end{bmatrix} 
    = \begin{bmatrix}
        (\bs{1}_{n_1}\bs{\rho}_1(\mathcal{D}_3)^T)\odot \begin{bmatrix}
            \bs{0}_{(n_1-n_3)\times n_3}\\
            \bs{I}_{n_3}
        \end{bmatrix}\\
        (\bs{1}_{n_2}\bs{\rho}_2(\mathcal{D}_3)^T)\odot \begin{bmatrix}
            \bs{0}_{(n_2-n_3)\times n_3}\\
            \bs{I}_{n_3}
        \end{bmatrix}
    \end{bmatrix}.
\end{align*}
We then apply block matrix inversion to simplify $\bs{V}_3^{-1}$:
\begin{align*}
    \bs{V}_3^{-1} =\begin{bmatrix}
        \bs{V}_2^{-1}+\bs{W}_{11} & \bs{W}_{12}\\
        \bs{W}_{12}^T & \bs{W}_{22}
    \end{bmatrix}
    = \begin{bmatrix}
        \bs{V}_2^{-1}+\bs{V}_2^{-1}\bs{U}_2\bs{Q}_3^{-1}\bs{U}_2^T\bs{V}_2^{-1} & -\bs{V}_2^{-1}\bs{U}_2\bs{Q}_3^{-1} \\
        -\bs{Q}_3^{-1}\bs{U}_2^T\bs{V}_2^{-1} & \bs{Q}_3^{-1}
    \end{bmatrix},
\end{align*}
where 
\small
\begin{align*}
    \quad \bs{V}_2^{-1} &= \begin{bmatrix}
        (\bs{V}^{(1,1)})^{-1} & \bs{0}\\
        \bs{0} & (\bs{V}^{(2,2)})^{-1}
    \end{bmatrix},\\
    \\
    \bs{W}_{22} &= \bs{Q}_3^{-1} = (\bs{V}^{(3,3)} - \bs{U}_2^T\bs{V}_2^{-1}\bs{U}_2)^{-1} 
    = \frac{1}{\sigma_3^2}\cdot \bs{R}_3(\mathcal{D}_3)^{-1},\\
    \\
    \bs{W}_{12} &= -\bs{V}_2^{-1}\bs{U}_2\bs{Q}_3^{-1} =
    -\begin{bmatrix}
            \bs{0}_{(n_1-n_3)\times n_3}\\
            \frac{1}{\sigma_3^2}(\bs{\rho}_1(\mathcal{D}_3)\bs{1}_{n_3}^T)\odot \bs{R}_3(\mathcal{D}_3)^{-1}\\
            \bs{0}_{(n_2-n_3)\times n_3}\\
            \frac{1}{\sigma_3^2}(\bs{\rho}_2(\mathcal{D}_3)\bs{1}_{n_3}^T)\odot \bs{R}_3(\mathcal{D}_3)^{-1}
    \end{bmatrix},
\end{align*}

\begin{align*}
    \bs{W}_{11} &= \bs{V}_2^{-1}\bs{U}_2\bs{Q}_3^{-1}\bs{U}_2^T\bs{V}_2^{-1}\\
    &= \Scale[0.95]{\begin{bmatrix}
        \bs{0}_{(n_1-n_3)\times(n_1-n_3)} & \bs{0}_{(n_1-n_3)\times n_3} & \bs{0}_{(n_1-n_3)\times(n_2-n_3)} & \bs{0}_{(n_1-n_3)\times n_3}\\
        \bs{0}_{n_3\times(n_1-n_3)} & \frac{1}{\sigma_3^2}(\bs{\rho}_1(\mathcal{D}_3)\bs{\rho}_1(\mathcal{D}_3)^T)\odot \bs{R}_3(\mathcal{D}_3)^{-1} & \bs{0}_{n_3\times(n_2-n_3)} & \frac{1}{\sigma_3^2}(\bs{\rho}_1(\mathcal{D}_3)\bs{\rho}_2(\mathcal{D}_3)^T)\odot \bs{R}_3(\mathcal{D}_3)^{-1}\\
        \bs{0}_{(n_2-n_3)\times(n_1-n_3)} & \bs{0}_{(n_2-n_3)\times n_3} & \bs{0}_{(n_2-n_3)\times(n_2-n_3)} & \bs{0}_{(n_2-n_3)\times n_3}\\
        \bs{0}_{n_3\times(n_1-n_3)} & \frac{1}{\sigma_3^2}(\bs{\rho}_2(\mathcal{D}_3)\bs{\rho}_1(\mathcal{D}_3)^T)\odot \bs{R}_3(\mathcal{D}_3)^{-1} & \bs{0}_{n_3\times(n_2-n_3)} & \frac{1}{\sigma_3^2}(\bs{\rho}_2(\mathcal{D}_3)\bs{\rho}_2(\mathcal{D}_3)^T)\odot \bs{R}_3(\mathcal{D}_3)^{-1}\\
    \end{bmatrix}}.
\end{align*}
\normalsize

\sloppy Next, we simplify the expression for $\bs{v}_3(\bs{x})^T\bs{V}_3^{-1}$, where $\bs{v}_3(\bs{x})$ is the covariance vector between $Z_3(\bs{x})$ and $\bs{z}^{(3)}=\{\bs{z}_1,\bs{z}_2,\bs{z}_3\}$ (observations at the three nodes). Then $\bs{v}_3(\bs{x}) = \left[\bs{v}_1^*(\bs{x},\mathcal{D}_1)^T,\bs{v}_2^*(\bs{x},\mathcal{D}_2)^T,\bs{v}_3^*(\bs{x},\mathcal{D}_3)^T\right]^T$, where

\begin{align*}
\begin{cases}
    \bs{v}_1^*(\bs{x},\mathcal{D}_1)^T &= \text{Cov}(Z_3(\bs{x}),Z_1(\mathcal{D}_1))^T = \rho_1(\bs{x})\cdot\text{Cov}(Z_1(\bs{x}),Z_1(\mathcal{D}_1))^T = \sigma_1^2\cdot \rho_1(\bs{x})\cdot \bs{r}_1(\bs{x},\mathcal{D}_1)^T\\
    \bs{v}_2^*(\bs{x},\mathcal{D}_2)^T &= \sigma_2^2\cdot \rho_2(\bs{x})\cdot \bs{r}_2(\bs{x},\mathcal{D}_2)^T\\
    \bs{v}_3^*(\bs{x},\mathcal{D}_3)^T &= \bs{\rho}_1(\mathcal{D}_3)\odot \bs{v}_1^*(\bs{x},\mathcal{D}_3)^T + \bs{\rho}_2(\mathcal{D}_3)\odot \bs{v}_2^*(\bs{x},\mathcal{D}_3)^T + \sigma_3^2\cdot \bs{r}_3(\bs{x},\mathcal{D}_3)^T.
\end{cases}
\end{align*}

We can thus rewrite $\bs{v}_3(\bs{x})^T\bs{V}_3^{-1} = \left[\bs{A}, \bs{B}\right]$ where
\begin{align*}
\begin{cases}
    \bs{A} &= \left[\rho_1(\bs{x})\cdot \bs{v}_1(\bs{x})^T(\bs{V}^{(1,1)})^{-1},\rho_2(\bs{x})\cdot \bs{v}_2(\bs{x})^T(\bs{V}^{(2,2)})^{-1}\right]\\ 
    &\quad - \left[\bs{0}_{n_1-n_3}, (\bs{\rho}_1(\bs{x})^T\odot \bs{r}_3(\bs{x},\mathcal{D}_3))\bs{R}_3(\mathcal{D}_3)^{-1},\bs{0}_{n_2-n_3}, (\bs{\rho}_2(\bs{x})^T\odot \bs{r}_3(\bs{x},\mathcal{D}_3))\bs{R}_3(\mathcal{D}_3)^{-1}\right]\\
    \bs{B} &= \bs{r}_3(\bs{x},\mathcal{D}_3)\bs{R}_3(\mathcal{D}_3)^{-1}.
\end{cases}
\end{align*}

Now, we show that the mean and covariance function of the GMGP model ((6) in main paper) match that for r-GMGP ((8) in main paper). Let $\bs{H}_3$ be the matrix of basis functions and $\bs{\beta}=\{\bs{\beta}_1,\bs{\beta}_2,\bs{\beta}_3\}$ be the vector of coefficients such that $\bs{H}_3\bs{\beta}$ gives the prior mean of $\bs{z}^{(3)}$. Then $\bs{H}_3$ is given by:

\begin{equation*}
    \bs{H}_3 = \begin{bmatrix}
        \bs{h}_1(\mathcal{D}_1)^T & \bs{0} & \bs{0}\\
        \bs{0} & \bs{h}_2(\mathcal{D}_2)^T & \bs{0}\\
        \bs{\rho}_1(\mathcal{D}_3)\odot \bs{h}_1(\mathcal{D}_3)^T & \bs{\rho}_2(\mathcal{D}_3)\odot \bs{h}_2(\mathcal{D}_3)^T & \bs{h}_3(\mathcal{D}_3)^T
    \end{bmatrix}.
\end{equation*}
Under a similar definition, we have $\bs{H}_1 = \bs{h}_1(\mathcal{D}_1)^T$ and $\bs{H}_2=\bs{h}_2(\mathcal{D}_2)^T$ such that the prior means of $\bs{z}_1$ and $\bs{z}_2$ are given by $\bs{H}_1\bs{\beta}_1$ and $\bs{H}_2\bs{\beta}_2$, respectively. Using the above formula for $\bs{v}_3(\bs{x})^T\bs{V}_3^{-1}$, we thus have
\begin{align*}
    \bs{v}_3(\bs{x})^T\bs{V}_3^{-1}\bs{z}^{(3)} &= \rho_1(\bs{x})\cdot \bs{v}_1(\bs{x})^T(\bs{V}^{(1,1)})^{-1}\bs{z}_1 + \rho_2(\bs{x})\cdot \bs{v}_2(\bs{x})^T(\bs{V}^{(2,2)})^{-1}\bs{z}_2\\
    &\quad - (\bs{\rho}_1(\mathcal{D}_3)\odot \bs{r}_3(\bs{x},\mathcal{D}_3))\bs{R}_3(\mathcal{D}_3)^{-1}\bs{z}_1(\mathcal{D}_3) - (\bs{\rho}_2(\mathcal{D}_3)\odot \bs{r}_3(\bs{x},\mathcal{D}_3))\bs{R}_3(\mathcal{D}_3)^{-1}\bs{z}_2(\mathcal{D}_3)\\
    &\quad + \bs{r}_3(\bs{x},\mathcal{D}_3)\bs{R}_3(\mathcal{D}_3)^{-1}\bs{z}_3,\\
    \bs{v}_3(\bs{x})^T\bs{V}_3^{-1}\bs{H}_3\bs{\beta} &= \rho_1(\bs{x})\cdot \bs{v}_1(\bs{x})^T(\bs{V}^{(1,1)})^{-1}\bs{H}_1\bs{\beta}_1 + \rho_2(\bs{x})\cdot \bs{v}_2(\bs{x})^T(\bs{V}^{(2,2)})^{-1}\bs{H}_2\bs{\beta}_2\\
    &\quad + \bs{r}_3(\bs{x},\mathcal{D}_3)\bs{R}_3(\mathcal{D}_3)^{-1}\bs{h}_3(\mathcal{D}_3)^T\bs{\beta}_3,\\
    \bs{v}_3(\bs{x})^T\bs{V}_3^{-1}\bs{v}_3(\cbl{\bs{x}'}) &= \cbl{\rho_1(\bs{x})\rho_1(\bs{x}')}\bs{v}_1(\bs{x})^T(\bs{V}^{(1,1)})^{-1}\bs{v}_1(\cbl{\bs{x}'}) + \cbl{\rho_2(\bs{x})\rho_2(\bs{x}')}\bs{v}_2(\bs{x})^T(\bs{V}^{(2,2)})^{-1}\bs{v}_2(\cbl{\bs{x}'})\\
    &\quad + \sigma_3^2\bs{r}_3(\bs{x},\mathcal{D}_3)\bs{R}_3(\mathcal{D}_3)^{-1}\bs{r}_3(\cbl{\bs{x}'},\mathcal{D}_3)^T.
\end{align*}

\cbl{We then derive the means and variances for the r-GMGP model (equations (8) in the main paper).}

\begin{align*}
    \mu_{Z_3}(\bs{x}) &=  \left[\sum_{t'\in Pa(3)}\rho_{t'}(\bs{x})\bs{h}_{t'}(\bs{x})^T\bs{\beta}_{t'}+\bs{h}_3(\bs{x})^T\bs{\beta}_3\right] + \bs{v}_3(\bs{x})^T\bs{V}_3^{-1}(\bs{z}^{(3)}-\bs{H}_3\bs{\beta})\\
    &= \rho_{1}(\bs{x})\bs{h}_{1}(\bs{x})^T\bs{\beta}_{1} + \rho_{2}(\bs{x})\bs{h}_{2}(\bs{x})^T\bs{\beta}_{2} + \bs{h}_3(\bs{x})^T\bs{\beta}_3\\
    &\quad + \rho_1(\bs{x}) \bs{v}_1(\bs{x})^T(\bs{V}^{(1,1)})^{-1}(\bs{z}_1-\bs{H}_1\bs{\beta}_1) + \rho_2(\bs{x}) \bs{v}_2(\bs{x})^T(\bs{V}^{(2,2)})^{-1}(\bs{z}_2-\bs{H}_2\bs{\beta}_2)\\
    &\quad + \bs{r}_3(\bs{x},\mathcal{D}_3)\bs{R}_3(\mathcal{D}_3)^{-1}\left[\bs{z}_3 - \bs{\rho}_1(\mathcal{D}_3)\odot \bs{z}_1(\mathcal{D}_3) - \bs{\rho}_2(\mathcal{D}_3)\odot \bs{z}_2(\mathcal{D}_3) - \bs{h}_3(\mathcal{D}_3)^T\bs{\beta}_3\right]\\
    &= \left[\rho_1(\bs{x})\mu_{Z_1}(\bs{x}) + \rho_2(\bs{x})\mu_{Z_2}(\bs{x}) + \bs{h}_3(\bs{x})^T\bs{\beta}_3\right]\\
    &\quad + \bs{r}_3(\bs{x},\mathcal{D}_3)\bs{R}_3(\mathcal{D}_3)^{-1}\left[\bs{z}_3 - \bs{\rho}_1(\mathcal{D}_3)\odot \bs{z}_1(\mathcal{D}_3) - \bs{\rho}_2(\mathcal{D}_3)\odot \bs{z}_2(\mathcal{D}_3) - \bs{h}_3(\mathcal{D}_3)^T\bs{\beta}_3\right],\\
    \\
    \sigma_{Z_3}^2(\bs{x},\cbl{\bs{x}'}) &= v^2_{Z_3}(\bs{x},\cbl{\bs{x}'}) - \bs{v}_3(\bs{x})^T\bs{V}_3^{-1}\bs{v}_3(\cbl{\bs{x}'})\\
    &= \cbl{\rho_1(\bs{x})\rho_1(\bs{x}')}\sigma_1^2 + \cbl{\rho_2(\bs{x})\rho_2(\bs{x}')}\sigma_2^2 + \sigma_3^2 -\cbl{\rho_1(\bs{x})\rho_1(\bs{x}')}\bs{v}_1(\bs{x})^T(\bs{V}^{(1,1)})^{-1}\bs{v}_1(\cbl{\bs{x}'})\\
    &-\cbl{\rho_2(\bs{x})\rho_2(\bs{x}')}\bs{v}_2(\bs{x})^T(\bs{V}^{(2,2)})^{-1}\bs{v}_2(\cbl{\bs{x}'}) - \sigma_3^2\bs{r}_3(\bs{x},\mathcal{D}_3)\bs{R}_3(\mathcal{D}_3)^{-1}\bs{r}_3(\cbl{\bs{x}'},\mathcal{D}_3)^T\\
    &= \cbl{\rho_1(\bs{x})\rho_1(\bs{x}')}\sigma^2_{Z_1}(\bs{x},\cbl{\bs{x}'}) + \cbl{\rho_2(\bs{x})\rho_2(\bs{x}')}\sigma^2_{Z_2}(\bs{x},\cbl{\bs{x}'}) + \sigma_3^2\left[1-\bs{r}_3(\bs{x},\mathcal{D}_3)^T\bs{R}_3(\mathcal{D}_3)^{-1}\bs{r}_3(\cbl{\bs{x}'},\mathcal{D}_3)\right].
\end{align*}

\cbl{With this, we can now show the equivalence of predictive means and variances \cbl{(by letting $\bs{x}=\bs{x}'$)} for the GMGP and r-GMGP models (from equations (6) and (8) in the main paper):
\begin{align*}
    \mu_{Z_3}(\bs{x}) &=   \left[\rho_1(\bs{x})\mu_{Z_1}(\bs{x}) + \rho_2(\bs{x})\mu_{Z_2}(\bs{x}) + \bs{h}_3(\bs{x})^T\bs{\beta}_3\right]\\
    &\quad + \bs{r}_3(\bs{x},\mathcal{D}_3)\bs{R}_3(\mathcal{D}_3)^{-1}\left[\bs{z}_3 - \bs{\rho}_1(\mathcal{D}_3)\odot \bs{z}_1(\mathcal{D}_3) - \bs{\rho}_2(\mathcal{D}_3)\odot \bs{z}_2(\mathcal{D}_3) - \bs{h}_3(\mathcal{D}_3)^T\bs{\beta}_3\right],\\
    \\
    \sigma_{Z_3}^2(\bs{x}) &= \rho_1^2(\bs{x})\sigma^2_{Z_1}(\bs{x}) + \rho_2^2(\bs{x})\sigma^2_{Z_2}(\bs{x}) + \sigma_3^2\left[1-\bs{r}_3(\bs{x},\mathcal{D}_3)^T\bs{R}_3(\mathcal{D}_3)^{-1}\bs{r}_3(\bs{x},\mathcal{D}_3)\right].
\end{align*}}
\noindent As shown above, the same recursive relation holds between $\mu_{Z_3}(\bs{x})$ and $m_{Z_3}(\bs{x})$, and between  $\sigma^2_{Z_3}(\bs{x})$ and $s^2_{Z_3}(\bs{x})$ when the training data are nested and noise-free. Thus we have $\mu_{Z_3}(\bs{x})=m_{Z_3}(\bs{x})$ and $\sigma^2_{Z_3}(\bs{x})=s^2_{Z_3}(\bs{x})$. With the assumption that all source nodes are independent, one can extend the proof approach in a straight-forward manner for in-trees of depth 2 with more than two source nodes.

\subsection{Connecting sub-graphs}\label{Sec1.2.2}

Now, let us consider a more complicated DAG which connects several lower-level in-trees. Again, we show the mean and variance equivalence using the simplest case and extend it to in-trees with more branches. Let node $A$ and $B$ be the root nodes of two lower-level in-trees, which we will call them Tree $A$ and Tree $B$, respectively. Node $A,B,C$ forms a larger in-tree and node $C$ is the child node. Fig. \ref{fig:APP_connectedGraph} (left) shows connections between Tree $A$ and Node $C$, Tree $B$ and Node $C$.

\begin{figure}[H]
\begin{minipage}{0.35\linewidth}
    \centering
    \tikz{
     \node[latent] (A) at (0,0) {$A$};
     \node[latent] (B) at (4,0) {$B$};
     \node[latent] (C) at (2,-1.5) {$C$};
     \plate [inner sep=.3cm,xshift=.2cm,yshift=.7cm] {plate1} {(A)} {Tree $A$}; 
     \plate [inner sep=.3cm,xshift=.2cm,yshift=.7cm] {plate1} {(B)} {Tree $B$}; 
     \edge {A,B} {C}  }
\end{minipage}
\hspace{0.05\linewidth}
\begin{minipage}{0.6\linewidth}
    \centering
    \tikz{
     \node[latent] (A) at (0,0) {$A$};
     \node[latent] (B) at (2.5,0) {$B$};
     \node[latent] (M) at (6,0) {$M$};
     \node[latent] (T) at (3,-1.5) {$T$};
     \plate [inner sep=.3cm,xshift=.2cm,yshift=.7cm] {plate1} {(A)} {Tree $A$}; 
     \plate [inner sep=.3cm,xshift=.2cm,yshift=.7cm] {plate1} {(B)} {Tree $B$}; 
     \plate [inner sep=.3cm,xshift=.2cm,yshift=.7cm] {plate1} {(M)} {Tree $M$}; 
     \edge {A,B,M} {T};
     \draw[loosely dotted] (B) to (M)}
\end{minipage}
\caption{In-trees constructed by: (\textbf{left}) connecting $2$ lower-level in-trees; (\textbf{right}) connecting $M > 2$ lower-level in-trees.}
\label{fig:APP_connectedGraph} 
\end{figure}


Again, we reorder the nested design sets such that $\mathcal{D}_A=[\mathcal{D}_A\setminus \mathcal{D}_C,\mathcal{D}_C]$, $\mathcal{D}_B=[\mathcal{D}_B\setminus \mathcal{D}_C,\mathcal{D}_C]$. With the independence assumption between Tree $A$ and Tree $B$, we have the following full covariance matrix:

\begin{align*}
\bs{V}_C = 
    \begin{bmatrix}
        \bs{V}_A & \bs{0} & \bs{U}^{(A,C)}\\
        \bs{0} & \bs{V}_B & \bs{U}^{(B,C)}\\
        \bs{U}^{(A,C)^T} & \bs{U}^{(B,C)^T} & \bs{V}^{(C,C)}
    \end{bmatrix},
\end{align*}
where $\bs{V}_A$ and $\bs{V}_B$ are covariance matrices of Tree $A$ and $B$, respectively. Let $\{Z_{A_i}\}_{A_i\in \text{Anc}(A)}$ be the non-root nodes in Tree $A$ (with $A$ being the root node), we have

\begin{align*}
\bs{U}^{(A,C)} = 
    \begin{bmatrix}
        \text{Cov}(Z_{A_1}(\mathcal{D}_{A_1}), Z_C(\mathcal{D}_C))\\
        \text{Cov}(Z_{A_2}(\mathcal{D}_{A_2}), Z_C(\mathcal{D}_C))\\
        \cdots\\
        \text{Cov}(Z_{A}(\mathcal{D}_{A}), Z_C(\mathcal{D}_C))\\
    \end{bmatrix},
\end{align*}

where
\begin{align*}
    \text{Cov}(Z_{A_i}(\mathcal{D}_{A_i}),Z_C(\mathcal{D}_C)) &= \text{Cov}(Z_{A_i}(\mathcal{D}_{A_i}),\bs{\rho}_A(\mathcal{D}_C)\odot Z_A(\mathcal{D}_C)+\bs{\rho}_B(\mathcal{D}_C)\odot Z_B(\mathcal{D}_C)+\delta_C(\mathcal{D}_C))\\
    &= \text{Cov}(Z_{A_i}(\mathcal{D}_{A_i}),\bs{\rho}_A(\mathcal{D}_C)\odot Z_A(\mathcal{D}_C))\\
    &= (\bs{1}_{n_{A_i}}\bs{\rho}_A(\mathcal{D}_C)^T)\odot\text{Cov}(Z_{A_i}(\mathcal{D}_{A_i}),Z_A(\mathcal{D}_C))\\
    &= (\bs{1}_{n_{A_i}}\bs{\rho}_A(\mathcal{D}_C)^T)\odot \bs{V}^{(A_i,A)}(\mathcal{D}_{A_i},\mathcal{D}_C).
\end{align*}
Thus,

\begin{align*}
    \bs{U}^{(A,C)} &= (\bs{1}_{(\sum_{A_i\in \text{Anc}(A)}n_{A_i}+n_A)}\bs{\rho}_A(\mathcal{D}_C)^T)\odot
    \begin{bmatrix}
        \bs{V}^{(A_1,A)}(\mathcal{D}_{A_1},\mathcal{D}_C)\\
        \bs{V}^{(A_2,A)}(\mathcal{D}_{A_2},\mathcal{D}_C)\\
        \cdots\\
        \bs{V}^{(A,A)}(\mathcal{D}_{A},\mathcal{D}_C)
    \end{bmatrix}\\
    &= (\bs{1}_{(\sum_{A_i\in \text{Anc}(A)}n_{A_i}+n_A)}\bs{\rho}_A(\mathcal{D}_C)^T)\odot \bs{V}_A^{n_C},
\end{align*}
where $\bs{V}_A^{n_C}$ refers to the last $n_C$ columns of $\bs{V}_A$. With the same calculation for $\bs{U}^{(B,C)}$, we have
\begin{align*}
    \begin{bmatrix}
        \bs{V}_A & \bs{0}\\
        \bs{0} & \bs{V}_B
    \end{bmatrix}^{-1}
    \begin{bmatrix}
        \bs{U}^{(A,C)}\\
        \bs{U}^{(B,C)}
    \end{bmatrix}
    &=\begin{bmatrix}
        (\bs{1}_{(\sum_{A_i\in \text{Anc}(A)}n_{A_i}+n_A)}\bs{\rho}_A(\mathcal{D}_C)^T)\odot \begin{bmatrix}
            \bs{0}_{((\sum_{A_i\in \text{Anc}(A)}n_{A_i}+n_A)-n_C)\times n_C}\\
            \bs{I}_{n_C}
        \end{bmatrix}\\
        (\bs{1}_{(\sum_{B_j\in \text{Anc}(B)}n_{B_j}+n_B)}\bs{\rho}_B(\mathcal{D}_C)^T)\odot \begin{bmatrix}
            \bs{0}_{((\sum_{B_j\in \text{Anc}(B)}n_{B_j}+n_B)-n_C)\times n_C}\\
            \bs{I}_{n_C}
        \end{bmatrix}
        \end{bmatrix}.
\end{align*}

We observe that the above expression is comparable to $\bs{V}_2^{-1}\bs{U}_2$ defined in previous section. Similarly, using the same definition of matrix $\bs{Q}_C$, we see that it is comparable to $\bs{Q}_3$: $\bs{Q}_C = \text{Cov}(Z_C(\mathcal{D}_C)|Tree\ A, Tree\ B) = \sigma_C^2\bs{R}_C(\mathcal{D}_C)$.

Following the same derivation as in previous section, we are able to show $\mu_{Z_C}(\bs{x})=m_{Z_C}(\bs{x})$ and $\sigma_{Z_C}^2(\bs{x})=s_{Z_C}^2(\bs{x})$. The above proof can also be naturally generalized to a more complicated in-tree where we have multiple lower-level in-trees connecting to a new root node. Finally, note that all directed in-trees can be recursively constructed by connecting lower-level in-trees to a new root node. This completes the proof of the equivalence for the posterior mean and variance of GMGP and r-GMGP when the multi-fidelity DAG $\mathcal{G}$ is a directed in-tree.

\newpage

\section{Algorithm for d-GMGP prediction}
Algorithm 1 outlines the steps for d-GMGP prediction.

\begin{algorithm}[H]
\caption{Computing the posterior predictive distribution for d-GMGP}\label{alg:d-GMGP}
\textbf{Input:} Training data $\{\mathcal{D}_t,\bs{z}_t\}$ over all nodes $t \in V$; Monte Carlo sample size $N$. \\
\textbf{Output:} Samples from the posterior predictive distribution for the highest-fidelity node $[Z_{T}(\bs{x})| \bs{z}^{(T)}]$.
\begin{algorithmic}[1]
\State Order the DAG nodes $t = 1, \cdots, T$ in terms of non-increasing depth (i.e., from leaf nodes to the root node).
\State \textbf{For} each node $t = 1, \cdots, T$:\\
\hskip2em \textbf{If} $t\in V_S$ (i.e., if $t$ is a source node):\\
\hskip4em Fit a GP model on $Z_t\sim \mathcal{GP}(0,\sigma_t^2r_t(\bs{x},\bs{x'}))$ using data $\{\mathcal{D}_t,\bs{z}_t\}$.\\
\hskip2em \textbf{Else}:\\
\hskip4em Compute the posterior means $\mathbb{E}[Z_{t'}| \{\bs{z}_m\}_{m \in \text{Anc}(t')},\bs{z}_{t'}]$ at design set $\mathcal{D}_t$ for parent nodes $t'\in \text{Pa}(t)$.\\
\hskip4em Fit a GP model (11) on $Z_t$ with kernel (12) in the main paper.

\State \hskip2em \textbf{End If}
\State \textbf{End For}

\State \textbf{For} each node $t=1, \cdots, T$:\\
\hskip2em \textbf{If} $t\in V_S$ (i.e., if $t$ is a source node):\\
\hskip4em Draw $N$ samples $\{z^*_{t,(i)}\}_{i=1}^N$ from the posterior distribution $[Z_t(\bs{x})|\bs{z}_t]$.\\
\hskip2em \textbf{Else}:\\
\hskip4em For $i =1, \cdots, N$, draw samples $z^*_{t,(i)} \sim [Z_t(\bs{x})|\bs{z}_{t},\{z^*_{t',(i)}\}_{t'\in\text{Pa}(t)}]$.
\State \hskip2em \textbf{End If}
\State \textbf{End For}
\State \textbf{Return}: Monte Carlo samples $\{z^*_{T,(i)}\}_{i=1}^N$ from posterior distribution $[Z_T(\bs{x})|\bs{z}^{(T)}]$.
\end{algorithmic}
\end{algorithm}
\cbl{Note that the two ``for" loops can be combined into one with the functionality of both fitting the model and making predictions.}

\newpage

\section{5-dimensional experiment}\label{sec:5dexp}

In this section, we compare the performance of the different methods on a $d=5$-dimensional experiment, with the same 3-node graph as in the $d=1$-dimensional experiment. The high-fidelity function $H$ (taken from \cite{Friedman1991}) and the two low-fidelity functions $L_1$ and $L_2$ are given as follows:
\begin{equation*}
\begin{cases}
    Z_H(\bs{x}) = 10 \sin(\pi x_1 x_2)+20(x_3-0.5)^2 + 10 x_4 + 5 x_5\\
    Z_{L_1}(\bs{x}) = 10 \sin(4 x_1 x_2)+20(x_3-0.5)^2 + 10 x_4 + 5 (1.2 x_5)\\
    Z_{L_2}(\bs{x}) = 10 \sin(3 x_1 x_2)+20(0.8 x_3-0.5)^2 + 10 (x_4-0.1) + 5 x_5
\end{cases}
\end{equation*}
\noindent The design points for $L_1$, $L_2$ and $H$ are generated from a sliced Latin Hypercube Design (SLHD, \citealp{Ba2015SLHD}) with sample sizes 40, 40 and 10, respectively, and the points in $H$ are again nested within $L_1$ and $L_2$. $M=500$ random test samples are used for comparison.

\begin{figure}[!h]
    \centering
    \includegraphics[width=.99\linewidth]{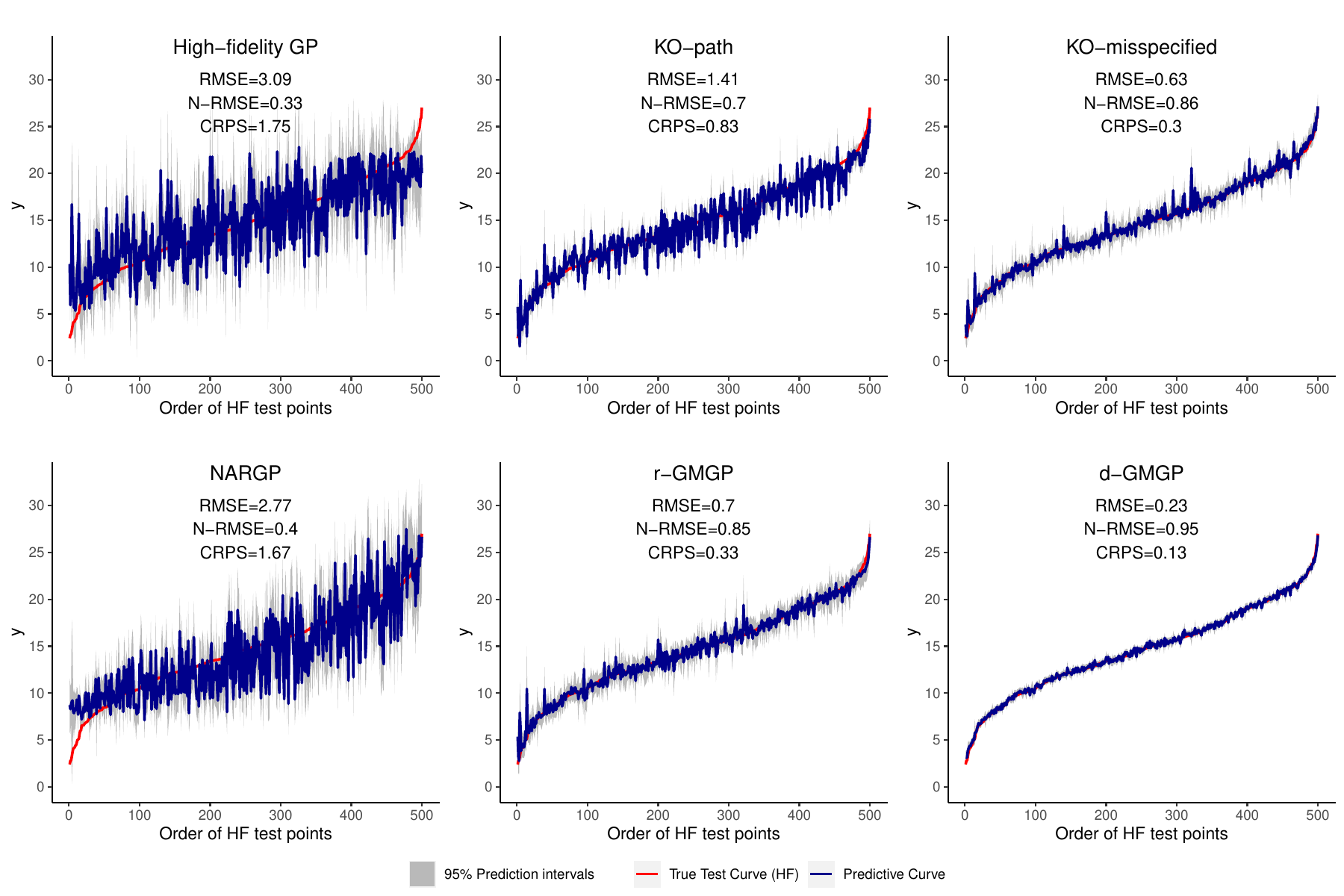}
  \caption{Results for the 5-d experiment: \cbl{three predictive metrics (RMSE, N-RMSE, CRPS)}, predictions (dark blue), 95\% predictive intervals (gray) and the true high-fidelity function (red). Here, the points are ordered from lowest to highest according to the value of high-fidelity test output.}
  \label{fig:Example5d}
\end{figure}

Fig. \ref{fig:Example5d} shows the predictive performance of the compared models. We see that, among the six plots, the predictive curve of the d-GMGP model is closest to the true test curve. The comparison of RMSE\cbl{, N-RMSE} and CRPS also shows that the proposed d-GMGP model indeed yields noticeably better predictive performance compared to other methods. This experiment again suggests that, when prior scientific information (in the form of a DAG) is available, integrating such information into a GP model can yield improved predictive performance. It is worth noting that the r-GMGP performs noticeably worse than the d-GMGP in this problem, and slightly worse than the KO-misspecified model. This is not unexpected, since again the true low and high-fidelity functions exhibit nonlinear correlations which are better captured by the d-GMGP model.

\section{20-dimensional experiment}\label{sec:5dexp}
\begin{figure}[!h]
    \centering
    \includegraphics[width=0.6\linewidth]{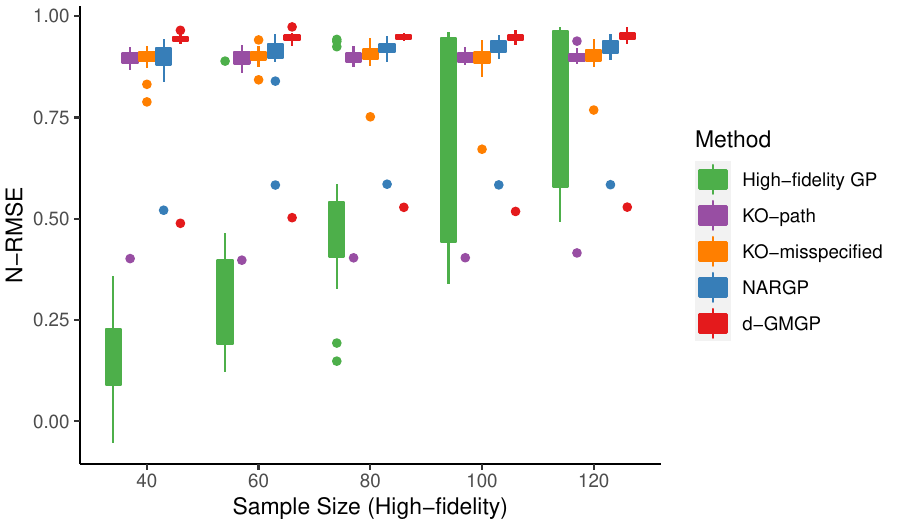}
    \caption{Results for the 20-d experiment: boxplots of N-RMSE for different sample sizes on $H$.}
    \label{fig:20d_NRMSE}
\end{figure}

\cbl{Figure \ref{fig:20d_NRMSE} shows the N-RMSE comparison for the 20-d experiment. Similar to RMSE and CRPS results, d-GMGP outperforms other methods with higher N-RMSE.}

\end{document}